%% file: article_aa29632_16.tex
\documentclass[traditabstract]{aa}

\usepackage{graphicx}
\usepackage{txfonts}
\usepackage{color}
\usepackage[switch, modulo]{lineno}
\usepackage{rotating}
\usepackage{xspace}
\usepackage{float}
\usepackage{url}

\usepackage{gensymb}
\usepackage{threeparttable}

\newcommand{\lt}{\textit{l}\xspace}
\newcommand{\bt}{\textit{b}\xspace}
\newcommand{\g}{$\gamma$\xspace}
\newcommand{\wco}{$W_{\rm{CO}}$\xspace}
\newcommand{\wcoavg}{$\overline{W_{\rm{CO}}}$\xspace}

\newcommand{\hd}{H$_2$\xspace}
\newcommand{\cosat}{$\rm{CO_{\rm sat}}$\xspace}
\newcommand{\hi}{{\sc H\,{\sc i} }\xspace}
\newcommand{\hii}{{H\,{\sc ii} }\xspace}
\newcommand{\nh}{$N_{\rm{H}}$\xspace}
\newcommand{\nhi}{$N_{\rm{HI}}$\xspace}
\newcommand{\nhii}{$N_{\rm{HII}}$\xspace}
\newcommand{\nhd}{$N_{\rm{H}_2}$\xspace}

\newcommand{\nhgam}{$N_{\rm{H}\,\gamma}$\xspace}
\newcommand{\nhlam}{$N_{\rm{H}\,m\lambda}$\xspace}
\newcommand{\Iff}{$I_{\rm{ff}}$\xspace}
\newcommand{\opa}{$\tau_{353}/N_{\rm{H}}$\xspace}
\newcommand{\opavghi}{$\overline{\tau_{353}/N}_{\rm{H}}^{\rm{HI}}$\xspace}
\newcommand{\opavgdnm}{$\overline{\tau_{353}/N}_{\rm{H}}^{\rm{DNM}}$\xspace}
\newcommand{\opavgco}{$\overline{\tau_{353}/N}_{\rm{H}}^{\rm{CO}}$\xspace}
\newcommand{\opaunit}{$10^{-27}$ cm$^2$ \xspace}
\newcommand{\av}{$A_{\rm{V}}$\xspace}
\newcommand{\avavgco}{$\overline{A_{\rm{V}}^{\rm CO}}$\xspace}
\newcommand{\sigvavg}{$\overline{\sigma_v^{\rm CO}}$\xspace}
\newcommand{\anh}{$A_{\rm{V}}/N_{\rm{H}}$\xspace}

\newcommand{\anhunit}{$10^{-22}$ mag. cm$^2$\xspace}
\newcommand{\qhi}{$q_{\rm{HI}}$\xspace}
\newcommand{\qco}{$q_{\rm{CO}}$\xspace}
\newcommand{\qdnm}{$q_{\rm{DNM}}$\xspace}
\newcommand{\qlis}{$q_{\rm{LIS}}$\xspace}

\newcommand{\xco}{$X_{\rm{CO}}$\xspace}
\newcommand{\xcoG}{$X_{\rm{CO} \gamma}$\xspace}

\newcommand{\xcounit}{$10^{20}$ cm$^{-2}$ K$^{-1}$ km$^{-1}$ s\xspace}
\newcommand{\taunu}{$\tau_{353}$\xspace}

\newcommand{\spw}{$4\pi R/N_{\rm{H}}$\xspace}

\newcommand{\nhebv}{N$_{\rm H}/E$(B-V)\xspace}
\newcommand{\msq}{$^{-2} $\xspace}

\begin{document}

   \title{Cosmic rays, gas and dust in nearby anticentre clouds: \\ I -- CO-to-\hd conversion factors and dust opacities}

\author{
Q.~Remy$^{(1)}$ \and 
I.~A.~Grenier$^{(1)}$ \and 
D.J.~Marshall$^{(1)}$ \and 
J.~M.~Casandjian$^{(1)}$ 
}
\authorrunning{LAT collaboration}

\institute{
\inst{1}~Laboratoire AIM, CEA-IRFU/CNRS/Universit\'e Paris Diderot, Service d'Astrophysique, CEA Saclay, F-91191 Gif sur Yvette, France\\ 
\email{quentin.remy@cea.fr} \\
\email{isabelle.grenier@cea.fr} \\
}

%
			
   \date{Received 1 September 2016 / Accepted 21 February 2017}
   
\abstract
{}
{We aim to explore the capabilities of dust emission and \g rays for probing the properties of the interstellar medium in the nearby anti-centre region, using 
 \g-ray observations with the \textit{Fermi} Large Area Telescope ({LAT}), and the thermal dust optical depth inferred from \textit{Planck} and \textit{IRAS} observations. 
We also aim to study massive star-forming clouds including 
the well known Taurus, Auriga, Perseus, and California molecular clouds, as well as a more diffuse structure which we refer to as Cetus.
In particular, we aim at quantifying potential variations in cosmic-ray density and dust properties per gas nucleon across the different gas phases and different clouds, and at measuring the CO-to-H$_2$ conversion factor, \xco, in different environments.
} 
{We have separated six nearby anti-centre clouds that are coherent in velocities and distances, from the Galactic-disc background in \hi 21-cm and $^{12}$CO 2.6-mm line emission. We have jointly modelled the \g-ray intensity recorded between 0.4 and 100 GeV, and the dust optical depth \taunu at 353 GHz  as a combination of \hi-bright, CO-bright, and ionised gas components. The complementary information from dust emission and \g rays was used to reveal the gas not seen, or poorly traced, by \hi, free-free, and $^{12}$CO emissions, namely (i) the opaque \hi and diffuse \hd present in the Dark Neutral Medium at the atomic-molecular transition, and (ii) the dense \hd to be added where $^{12}$CO lines saturate.   }
{The measured interstellar \g-ray spectra support a uniform penetration of the cosmic rays with energies above a few GeV through the clouds, from the atomic envelopes to the $^{12}$CO-bright cores, and with a small $\pm 9$\% cloud-to-cloud dispersion in particle flux. We detect the ionised gas from the \hii region NGC1499 in the dust and \g-ray emissions and measure its mean electron density and temperature. We find a gradual increase in grain opacity as the gas (atomic or molecular) becomes more dense. The increase reaches a factor of four to six in the cold molecular regions that are well shielded from stellar radiation. Consequently, the \xco factor derived from dust is systematically larger by 30\% to 130\% than the \g-ray estimate. We also evaluate the average \g-ray \xco factor for each cloud, and find that \xco tends to decrease from diffuse to more compact molecular clouds, as expected from theory. 
We find \xco factors in the anti-centre clouds close to or below \xcounit, in agreement with other estimates in the solar neighbourhood. Together, they confirm the long-standing unexplained discrepancy, by a factor of two, between the mean \xco values measured at parsec scales in nearby clouds and those obtained at kiloparsec scale in the Galaxy. Our results also highlight large quantitative discrepancies in $^{12}$CO intensities between simulations and observations at low molecular gas densities.
}
{}

\titlerunning{gas \& dust in nearby anticentre clouds I}

\maketitle

\section{Introduction}

The structural, dynamical and thermodynamical evolution of interstellar clouds is often probed with observations of the \hi 21-cm line for the atomic gas \citep{2005AeA...440..775K,2010AeA...521A..17K,2011ApJS..194...20P} and of CO rotational emission lines for the molecular gas \citep{2015ARAeA..53..583H}. However, measuring gas column densities in the dense media, both atomic and molecular, is hampered by the opacity of the interstellar medium (ISM) to these radiations and we have no efficient means to correct for line opacities in the absence of absorption data against background point sources. Complementary information can be gained from 
the sub-millimeter and infrared thermal emission of large dust grains mixed with the gas  \citep{2014AeA...571A..11P}, and from \g rays produced in interactions of high-energy cosmic rays (CRs) with interstellar gas nucleons \citep{1982AeA...107..390L}.  Dust grains and CRs trace all chemical and thermodynamical forms of the gas to large depths into the clouds \citep{2015AeA...582A..31A}, but they trace only integrals of the gas column densities, \nh, along sight lines, and they bear no kinematic information. To use them, we rely on critical assumptions, namely a uniform dust-to-gas mass ratio and uniform grain emissivity across a cloud and, likewise, a uniform CR flux through a cloud. Those assumptions still need to be tested in a variety of clouds and across complex phase changes, especially on consideration that dust grains and their radiation properties vary from the diffuse ISM to molecular clouds  \citep{2003AeA...398..551S,2012ApJ...751...28M,2013ApJ...763...55R,2013AeA...559A.133Y,2009ApJ...701.1450F,2014AeA...571A..11P,2014AeA...566A..55P,2015AeA...582A..31A}.  One can also test the uniformity of the \g-ray emissivity spectrum of the gas in well-sampled nearby clouds to verify the smooth penetration of CRs with energies above ${\sim}1$ GeV \citep{1976AeA....53..253S,1978AeA....70..367C,2015MNRAS.451L.100M}.

The recent progress in \hi, CO, dust, and \g-ray observations provides excellent opportunities to confront these observations at a resolution of a few parsecs inside nearby clouds to probe biases in the different tracers and to search for missing gas in the census. This comparison can shed light on the highly opaque atomic gas forming the cold neutral medium (CNM), where the \hi brightness temperature probes the excitation (spin) temperature of the gas rather than the column density \citep[][and references therein]{2015ApJ...804...89M}. 
One can also probe how the $^{12}$CO, $J{=}1{\rightarrow}$0, emission rapidly saturates at high column density and how it fades out in the diffuse molecular gas where CO lines are too weakly excited and/or CO molecules are too efficiently photodissociated to map the corresponding molecular hydrogen \citep{1988ApJ...334..771V,2010ApJ...716.1191W}. Large \hi opacities and  the lack of CO emission both conspire  to hide large gas masses at the H--\hd interface. This dark neutral medium (DNM) can be jointly revealed by cosmic rays and dust \citep{2005Sci...307.1292G}. 

A more precise census of the gas column densities is key to establishing and quantifying variations in dust opacity (optical depth per gas nucleon) across cloud structures. Mapping those variations as a function of grain temperature, $T$, and opacity spectral index, $\beta$ gives important diagnostics for grain evolution through mantle accretion, coagulation, and ice coating \citep{2015AeA...579A..15K}.


With an approach similar to the multi-tracer analysis of the Chamaeleon complex \citep{2015AeA...582A..31A}, we present an analysis of the anticentre region, which exhibits more massive clouds, with higher levels of star-formation activity, so we can test the properties of the gas tracers to larger column densities and molecular fractions. The analysed region spans a few hundred parsecs in distance in the local ISM (see below). It encompasses the well-known Taurus, Auriga, and Perseus molecular clouds \citep{1987ApJS...63..645U,2010AeA...512A..67L,2008ApJ...684.1228S}, the massive complex associated with the California nebula \citep{2009ApJ...703...52L}, and a nearby, diffuse, hardly molecular cloud, which we refer to as Cetus.

Our goal is to test the dust and \g-ray emission as tracers of the total gas against linear combinations of the emission line intensities recorded in the \hi and CO observations, a new free-free emission map of the ionised gas, plus the DNM gas which escapes the radio and millimeter spectroscopic surveys at the atomic-molecular interface. We have separated the \hi and CO clouds in space (i.e., direction, velocity, and distance) to study their individual contributions to the total dust and \g-ray signals. We have used the model results to derive scaling factors such as the CO-to-H${_2}$ conversion factor, \xco, in each cloud, and the average dust opacity in the different gas phases for each cloud. We have then used the multi-wavelength data to follow dust evolution with increasing \nh in several molecular clouds with masses spanning nearly two orders of magnitude. A companion paper will present the results on the transitions between gas phases, and on the mass fractions and contribution of CO-dark \hd to the molecular phase in individual regions.

The paper is structured as follows: the data are presented in Sect. 2 and the models are described in Sect. 3. The model results and the detection of gas in the DNM, CO-saturated, and ionised phases are detailed in Sect. 4. The CR content of the clouds, the \xco ratios, and the dust opacities are discussed in Sect. 5, 6, and 7, respectively. Additionally, we present the \hi and CO component separation into nearby cloud complexes in appendix \ref{sec:AnnexCS}, the derivation of the free-free intensities in appendix \ref{sec:AnnexFF}, the dependence of the goodness of fit on the \hi spin temperature in appendix \ref{sec:AnnexT}, the best-fit coefficients of the dust and $\gamma$-ray models in appendix \ref{sec:AnnexFITs}, and a table with historical \g-ray measurements of \xco factors in \ref{tab:Xco_history}.

\begin{figure*}[!ht]
\includegraphics[width=\hsize]{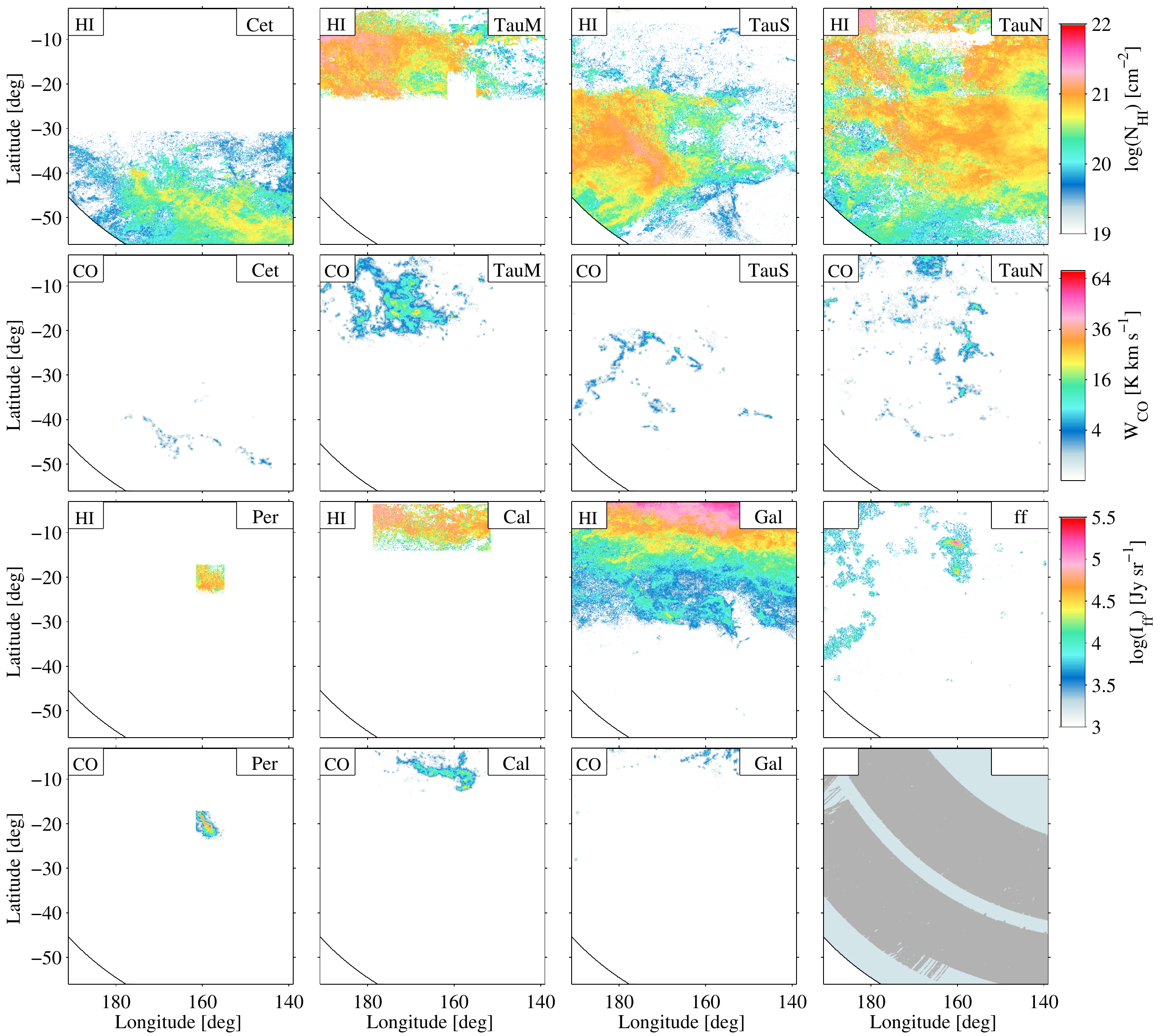}
\caption{Maps of the Cetus (Cet), Main Taurus (TauM), South Taurus (TauS), North Taurus (TauN), Perseus (Per), and California (Cal) clouds, and of the Galactic disc background (Gal) in \hi column density, \nhi, for a spin temperature of 400~K, and in CO intensity, \wco. The \nhi and \wco maps are respectively labelled HI and CO. The map labelled ff shows the intensity $I_{{\rm ff}}$ of free-free emission toward this region. The last subplot shows the coverage of the GALFA (dark grey) and EBHIS (light blue) \hi data.}  
\label{fig:NHI}
\end{figure*}

\section{Data}
 The analysis region extends from 139$^\circ$ to 191$^\circ$ in Galactic longitude and from $-56^\circ$ to $-3^\circ$ in Galactic latitude. 
We have masked regions with large column densities from the background Galactic disc, at $-9^\circ<b<-3^\circ$ for $l>183^\circ$ and $l<152^\circ$ and regions with small column densities at declinations below $-5$\degr where the \hi maps have lower angular resolution. All maps in Fig. \ref{fig:NHI} have been projected onto  the same 0\fdg125-spaced Cartesian grid as that of the CfA CO survey \citep{2001ApJ...547..792D}. We have checked that the use of the Cartesian projection, instead of equal-area bins, does not significantly bias the results. The largest differences are less than a fifth of the uncertainties found in the jackknife tests discussed in Sect. \ref{sec:jack}. 

\subsection{HI and CO data}\label{sec:HIdata}
 
For this analysis we have used the 4\arcmin-resolution GALFA-HI survey where available \citep{2011ApJS..194...20P} and we have complemented it with the 10.8\arcmin-resolution EBHIS survey elsewhere \citep{2016AeA...585A..41W}. As shown in Fig. \ref{fig:NHI}, the GALFA survey covers 74\% of the region. GALFA data-cubes \footnote{\url{https://purcell.ssl.berkeley.edu}} were re-sampled into the 0\fdg125-spaced Cartesian grid. We have used the narrow-band cubes with their original velocity resolution of 0.18 km/s in the local standard of rest (LSR). The EBHIS frequency sampling is coarser, with a velocity resolution of 1.44 km/s. In the analysed region covered by both survey we have found a tight correlation in column densities between the two:
\begin{equation}
N_{\rm{H}}^{\rm{GALFA}}=1.0013 \times N_{\rm{H}}^{\rm{EBHIS}} -8.17\times10^{17} \rm{cm}^{-2}.
\end{equation} 

In order to trace the molecular gas we have used $^{12}$CO $J{=}1{\rightarrow}0$ observations at 115 GHz from the moment-masked CfA CO survey of the Galactic plane \citep{2001ApJ...547..792D,2004ASPC..317...66D}. Most of the survey is based on a 0\fdg125-spaced Cartesian grid except for the high-latitude clouds, at $b\lesssim-50^\circ$, which have been interpolated from 0\fdg25 to 0\fdg125.

\subsection{HI and CO velocity separation}\label{sec:velsep}

We have decomposed the \hi and CO spectra into individual lines and we have used this information to identify six nearby entities that are coherent in velocity, position (in Galactic coordinates), and distance, in addition to the emission from the background Galactic disc. The \hi velocity resolution in the region covered by the EBHIS survey is coarser, but still adequate to separate the main structures because of the limited confusion along sight lines at medium-to-high latitudes. The separation process is three-fold and is described by \cite{2015AeA...582A..31A} and in Appendix \ref{sec:AnnexCS}. The choice of six main entities has been guided by the structure in (\lt, \bt, v) density and by distance information obtained from stellar reddening \citep{2014ApJ...786...29S}. Details on the (\lt, \bt, v) cuts are given in Table \ref{tab:CompLim}. The main structures found in the local ISM, namely Cetus, Main Taurus, South Taurus, North Taurus, Perseus, and California, together with the background Galactic structures are depicted in \nhi column densities and in \wco intensities in Fig. \ref{fig:NHI}. The set of \hi maps includes all of the \hi emission observed in this region except for the high velocity clouds \citep{2008ApJ...672..298W}. In order to investigate the effect of the unknown \hi optical depth, we have derived all the \nhi maps for a sample of uniform spin temperatures (150, 200, 300, 400, 500, 600, 700, and 800 K) and for the optically thin case. 

Distance estimates to the local clouds have been compiled from the photometric measurements of PanSTARRS-1 and the detection of reddening fronts toward stellar groups \citep{2014ApJ...786...29S}. We have selected the stars along lines of sight intersecting each (and only one) cloud within its \wco contour at 0.5 K km s$^{-1}$ and we have calculated the average distance in each sample. For the reddening fronts to the stars toward Cetus, Main Taurus, South Taurus, North Taurus, Perseus, and California we find average distances of 190$\pm$30~pc, 140$\pm$30~pc, 160$\pm$10~pc, 190$\pm$50~pc, 270$\pm$20~pc, and 410$\pm$20~pc, respectively. They compare reasonably with the geometry inferred from the modelling of the far-UV continuum emission \citep{2013ApJ...765..107L}. The North Taurus component is dominated by the Auriga cloud near the northern edge of our analysis region; it also includes MBM 12 molecular cloud from \cite{1985ApJ...295..402M} catalogue. The South Taurus component includes  MBM 6, MBM 12 and MBM 18.

\begin{figure*}
\includegraphics[width=\hsize]{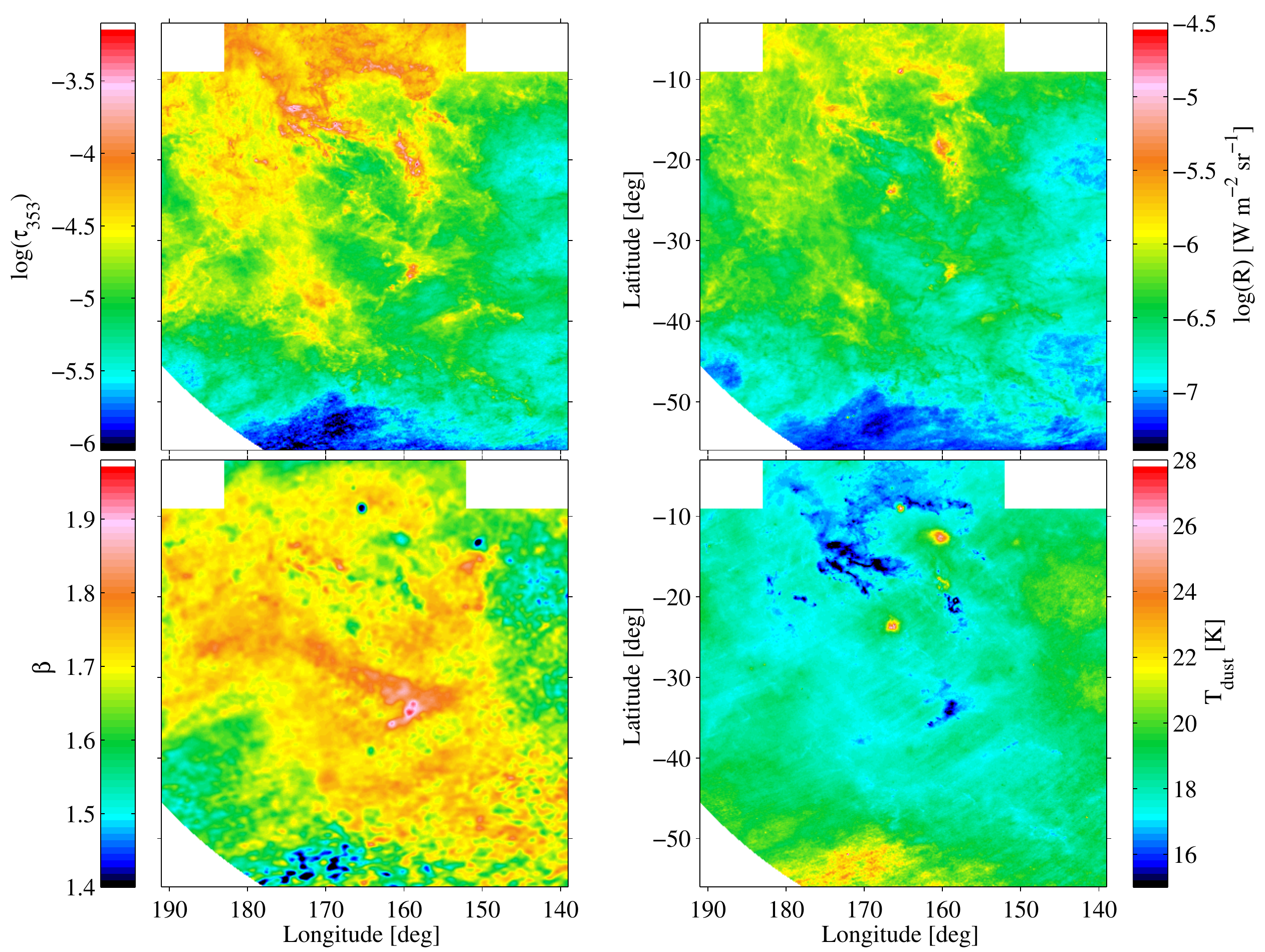}
\caption{Maps of dust optical depth at 353 GHz (top left), radiance (top right), spectral $\beta$ index (bottom left), and colour temperature (lower right), from \cite{2014AeA...571A..11P}.} 
\label{fig:PLK4}
\end{figure*}
 
\subsection{Dust data}

In order to trace the dust column density, we have used the optical depth, \taunu, obtained at 353 GHz by \cite{2014AeA...571A..11P}. They have modelled the spectral energy distributions of the thermal emission of the large grains with modified black-body curves in specific intensity, $I_{\nu}=\tau_{\nu_{0}} B_{\nu}(T_{\rm d}) (\nu /\nu_0)^{\beta}$.
The optical depth $\tau_{\nu_{0}}$ at frequency $\nu_0$, the dust colour temperature $T_{\rm d}$, and the spectral index $\beta$ were fitted to the data from \textit{Planck} and the Infrared Astronomical Satellite (IRAS)  across the sky at 30$\arcmin$ resolution. The fits were then repeated at 5$\arcmin$ resolution by fixing $\beta$ as obtained in the first step. This procedure limited the noise impact on the $T_{\rm d}$--$\beta$ degeneracy in faint regions. The modelled intensities were checked to be consistent with the data at all frequencies (see Fig. 11 in \citealt{2014AeA...571A..11P}). The contamination from CO line emission in the 353\,GHz filter band amounts to a few per cent of the total signal. It was not removed so as to avoid increasing the noise in directions away from CO clouds.

The optical depth, $\tau_{\nu}$, and opacity, $\sigma_{\nu}$, of the dust at frequency $\nu$ follow the relations :
\begin{equation}
\tau_{\nu}=\frac{I_{\nu}}{B_{\nu}(T_{\rm d})}=\sigma_{\nu}N_{H}.
\end{equation}
We have also calculated the total power radiated by the dust, also referred to as radiance: 
\begin{equation}
\mathcal{R}=\int \tau_{353} \, B_{\nu}(T_{\rm d})\left( \frac{\nu}{353\, {\rm GHz}} \right) ^{\beta} d\nu.
\end{equation}
Figure \ref{fig:PLK4} shows the dust properties thus measured across the analysis region. 

The corresponding analysis of the Chamaeleon region \citep{2015AeA...582A..31A} compared different dust tracers, namely the optical depth and an estimate of the dust extinction, $A_{VQ}$, empirically corrected for starlight intensity. The latter distribution best followed the gas column densities in the Chamaeleon region, but we have checked that this is not true in the anticentre region in particular toward regions of dust temperatures exceeding 23 K,, so we present the results of our analyses using only the \taunu optical depth. Understanding why the empirical correction used for $A_{VQ}$ is less efficient in the anticentre region will be investigated separately.


\subsection{Ionised gas}

Ionised gas is visible in H$\alpha$ emission \citep{2003ApJS..146..407F} in the California nebula around $l=160$\fdg1 and $b =-12$\fdg4, in the G159.6-18.5 \hii region, and along the Eridanus loop. The Califonia Nebula, alias NGC 1499 or S 220, is ionised by $\xi$ Persei, an O7.5III star located at a distance of $380 \pm 70$~pc \citep{2007AeA...474..653V} that has run away from the Perseus OB 2 association \citep{1976PASJ...28..437S}.
The 1\fdg2-diameter region G159.6-18.5 has been seen by IRAS and the Wide-Field Infrared Survey Explorer (WISE) \citep{2000AJ....119.1325A,2014ApJS..212....1A}. It is excited by another run-away O9.5V star, HD 278942, at a distance of $190 \pm 40$~pc \citep{2007AeA...474..653V}, which would place it in front rather than behind the Perseus cloud \citep{2008hsf1.book..308B}.

The composite H$\alpha$ map varies in angular resolution from 6\arcmin~to 1\degr~across the region of interest and H$\alpha$ emission can be efficiently absorbed behind dense clouds \citep[see for instance the dusty absorption lane across G159.6-18.5 in Fig. 8 of][]{2008hsf1.book..308B}. We have therefore preferred measures of the free-free intensity at mm wavelengths to trace the ionised gas. 

The \textit{Planck} LFI map at 70 GHz \citep[release 2.01 data,][]{2016AeA...594A...1P} shows free-free emission in spatial coincidence with the H$\alpha$ emission. The angular resolution of 14\arcmin~of LFI better suits our analysis than the 1\degr-resolution of the free-free map inferred at 22 GHz from the 9-year WMAP data \citep{2013ApJS..208...20B}. To separate the free-free intensity in the 70 GHz data, we have successively removed the contributions from the Cosmological Microwave Background (CMB), from dust emission, and from point sources. The method we have used to extract the free-free emission is described in Appendix \ref{sec:AnnexFF}.

The resulting free-free dominated map is displayed in Fig. \ref{fig:NHI}. It closely resembles the H$\alpha$ intensity map of the region \citep{2003ApJS..146..407F}, but for a uniform resolution of 14\arcmin~and a reduced contrast, by a factor 2 to 8, between the intensities of G159.6-18.5 and NGC 1499. because the Perseus clouds are more transparent to mm waves.

The integral of the free-free intensity along a sight line through ionised gas with temperature $T_e$ and electron and ion number densities, $n_e$ and $n_i$, scales at 70 GHz as :
\begin{equation}
I_{\rm ff} = 63.4 \;{\rm Jy \,sr^{-1}} \; \left(\frac{T_e}{10^4\, {\rm K}}\right)^{-0.28} \int n_e\,n_i\,ds,
\label{eq:Iff}
\end{equation}
where the emission measure $\int n_e\,n_i\,ds$ is expressed in cm$^{-6}$~pc \citep{1961ApJS....6..167K,1988AeA...189..319C}. If the observed intensity results from discrete and homogeneous nebulae with comparable uniform electron densities, $\overline{n}_e$, \Iff roughly scales with the ion column density,
\begin{equation}
$\nhii$ \approx 4.87\,10^{16}$~cm$^{-2} \;(I_{\rm ff}/{\rm 1\,Jy \,sr^{-1}})\;(T_e/10^4\, {\rm K})^{0.28} (\overline{n}_e/1\,{\rm cm^{-3}})^{-1}, 
\label{eq:NHII}
\end{equation}
so we have used the \Iff map as a template for the \nhii distribution in the gas model and we discuss the scaling ratio obtained in the fit between \nhii and \Iff in Sec. \ref{sec:ffres}.

\subsection{Gamma-ray data}

We have used six years of the Pass 8 photon data from \textit{Fermi}-LAT, the associated instrument response functions (IRFs, P8R2\_CLEAN\_V6), 
and the corresponding isotropic spectrum for the extragalactic and residual instrumental backgrounds \citep{2013arXiv1303.3514A,2015ApJS..218...23A}. We have applied tight rejection criteria (CLEAN class selection, photon arrival directions within < 100$^\circ$ of the Earth zenith and in time intervals when the LAT rocking angle was inferior to 52$^\circ$) in order to reduce the contamination by residual CRs and Earth atmospheric \g rays in the photon data \citep[see][for details]{2012ApJS..199...31N}. We have sampled the IRFs, the exposure map, the \g-ray emissivity spectrum, \qlis, of the local interstellar gas \citep{2015ApJ...806..240C}, and the spectrum of the isotropic background in 14 energy bins, 0.2 dex in width and centred from 10$^{2.3} $ to 10$^{4.9} $ MeV.

To ensure photon statistics robust enough to follow details in the spatial distributions of the different interstellar components, we have analysed the data in broad and independent energy bands, bounded by 
10$^{2.6} $, 10$^{2.8} $, 10$^{3.2} $, 10$^{3.6} $, and 10$^{5} $ MeV. We have also analysed the data in the integrated 10$^{2.6} $ -- 10$^{5} $ MeV band. The low-energy threshold was set to take advantage of the best LAT point-spread function (PSF), which strongly degrades at low energy. The analysis of the data in two lower-energy bands (10$ ^{2.2} $--10$ ^{2.4}$ and 10$ ^{2.4} $--10$ ^{2.6}$ MeV) proved to be less reliable to separate the gas phases in compact clouds. To optimise the angular resolution, we have selected all detected photons above 1.6~GeV, but only those at lower energy that converted to pairs in the front section of the tracker \citep{2009ApJ...697.1071A,2012ApJS..203....4A}. Given the broad ranges of the energy bands, we have not corrected the fluxes for the energy resolution, which is better than 14\% at these energies \citep{2013arXiv1303.3514A}. To account for the spill-over of emission produced outside the analysis region, but reconstructed inside it, we have modelled point sources and interstellar contributions in a region 4$^\circ$ wider than the analysis region. This choice corresponds to the 99.5$\%$ containment radius of the PSF in the lowest energy band. For the \hi emission outside the analysis region not covered by EBHIS and GALFA surveys we have used the Leiden/Argentine/Bonn (LAB) survey \citep{2005AeA...440..775K}.

The positions and the flux spectra of the \g-ray sources in the field are provided by the \textit{Fermi}-LAT Third Source Catalog \citep{2015ApJS..218...23A}. The observed $\gamma$-ray emission also includes a contribution
from the large-scale Galactic inverse Compton (IC) emission emanating from the interactions of CR electrons with the ISRF. The GALPROP\footnote{\url{http://galprop.stanford.edu}} parameter file 54-LRYusifovXCO4z6R30-Ts150-mag2 has been  used to generate an energy-dependent template of the Galactic IC emission across the analysis region \citep{2012ApJ...750....3A}.

\begin{figure}
  \centering
  \includegraphics[width=\hsize]{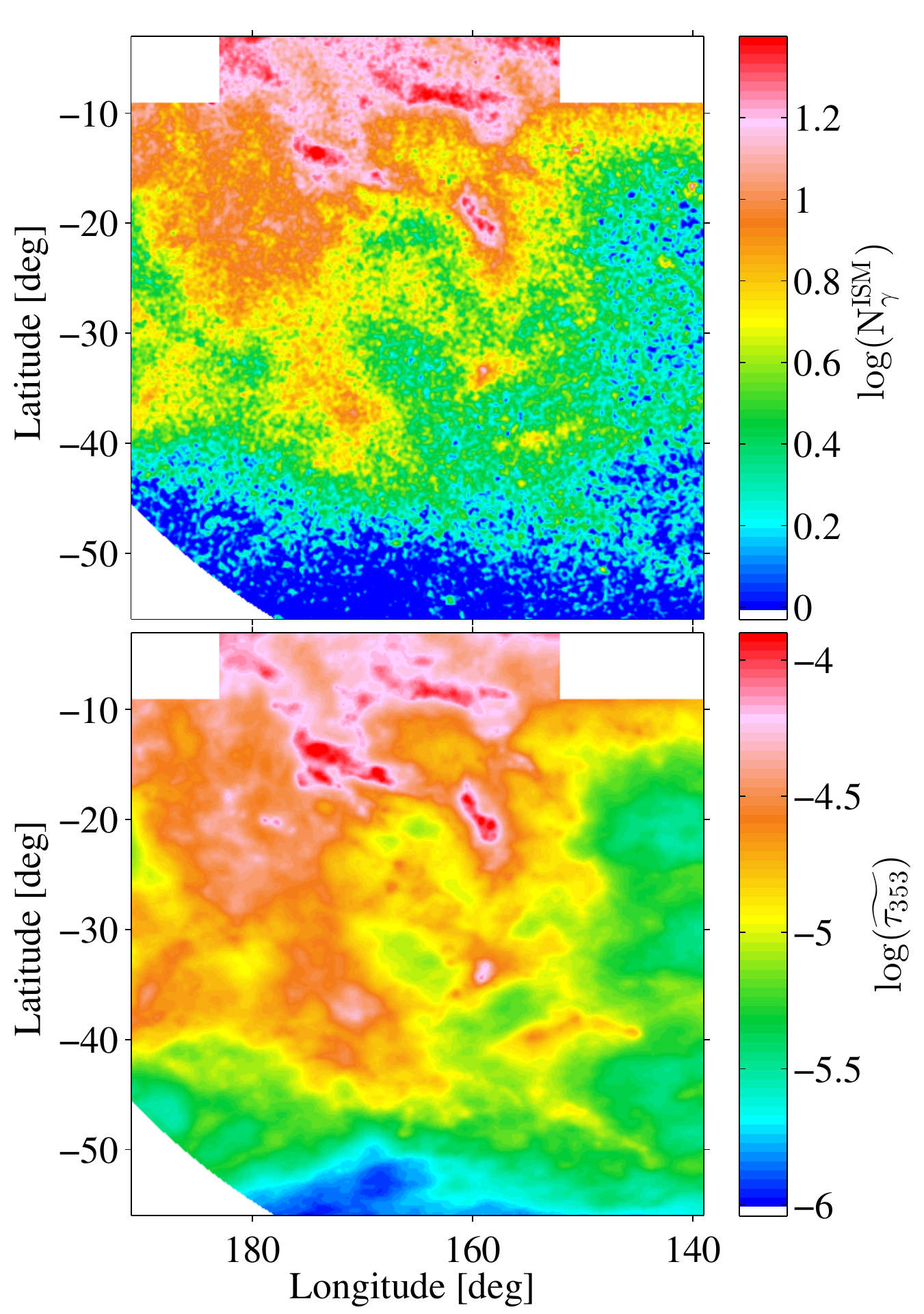}
  \caption{\textit{Top:} \g-ray counts of gaseous origin recorded in the 0.4--100 GeV energy band in a 0\fdg125 pixel grid. \g-ray emissions other than due to cosmic-ray interactions in the gas have been subtracted. The map has been smoothed with a Gaussian kernel of 0\fdg14 dispersion for display. \textit{Bottom:} dust optical depth measured at 353 GHz and displayed at the \textit{Fermi}-LAT angular resolution for comparison.}  
  \label{fig:gamDust}
\end{figure}

\section{Models and analyses}\label{sec:mod}

To first order, the dust and \g-ray emissions should both trace the total gas column density, \nh. In order to compare them in  Fig. \ref{fig:gamDust}, we have convolved the dust optical depth with the LAT PSF on the one hand, and we have subtracted non-gaseous \g-ray emissions obtained by the fitting (Sec. \ref{sec:results}) from the \g-ray data on the other hand. Figure \ref{fig:gamDust} shows strong similarities in the spatial distributions of both tracers, but it also reveals differences in their dynamical range in several places.

To detect neutral gas unaccounted for in \nhi and \wco, we have used the fact that it is permeated by both cosmic rays and dust grains. We have therefore extracted the significant \g-ray and dust residuals over the \nhi, \wco, and \nhii expectations and we have used the spatial correlation between those residuals to infer the additional gas column densities (see Sec \ref{sec:iter}). We have separated the residuals for two types of environments: in regions of weak or no CO intensity, below 7~K km s$^{-1}$, which correspond to the DNM at the H-H$_2$ transitions; and in regions toward large CO intensities, above 7~K km s$^{-1}$, to capture additional H$_2$ gas where $^{12}$CO emission saturates and rarer isotopologues such as $^{13}$CO or C$^{18}$O should be used. We refer to this saturated-CO molecular component as ``\cosat''.

\subsection{Dust model}

In the case of a uniform dust-to-gas mass ratio and uniform mass emission coefficient of the grains, the dust optical depth linearly scales with the total \nh. We have therefore modelled $\tau_{353}(l,b)$ in each direction as a linear combination of the gaseous contributions from the different phases (\hii, \hi, DNM, CO-bright, \cosat), with free normalisations to be fitted to the data, as in \citet{2015AeA...582A..31A}. We have added a free isotropic term, $y_{\rm{iso}}$, to account for the residual noise and the uncertainty in the zero level of the dust data \citep{2014AeA...571A..11P}. The $\tau_{353}(l,b)$ model can be expressed as:
\begin{eqnarray}
\tau_{353}(l,b)&=& \sum_{i=1}^7 y_{\rm{HI},i} N_{\rm{HI},i}(l,b) + \sum_{i=1}^7 y_{\rm{CO},i} W_{\rm{CO},i}(l,b) + y_{\rm{ff}}I_{\rm{ff}}(l,b) \notag\\
 &+& y_{\rm{DNM}} N_{\rm H}^{\rm{DNM}}(l,b) + y_{\rm{COsat}} N_{\rm H}^{\rm{COsat}}(l,b) + y_{\rm{iso}},
\label{eq:modDust}
\end{eqnarray} 
where $N_{\rm{HI},i}(l,b)$, $W_{\rm{CO},i}(l,b)$, and $I_{\rm{ff}}(l,b)$ respectively denote the \nhi, \wco, and free-free maps of the clouds depicted in Fig. \ref{fig:NHI}. $N_{\rm H}^{\rm{DNM}}(l,b)$ and $N_{\rm H}^{\rm{COsat}}(l,b) $ stand for the column densities in the DNM and \cosat components deduced from the coupled analyses of the \g-ray and dust data (see Sect. \ref{sec:iter}). 

The $y$ model parameters have been estimated using a $\chi^{2}$ minimization. We expect the model uncertainties to exceed the measurement errors in $\tau_{353}(l,b)$ because of potential variations in grain properties through the clouds and because of the limitations of the gas tracers (survey sensitivities, emission saturation, self-absorption, etc.). As we cannot precisely determine the model uncertainties, we have set them to a fractional value of the data and we have determined this fraction to be 19$\%$ by reaching a reduced $\chi^{2}$ of unity. 
This fraction is larger than the 3\% to 9\% error in the measurement of $\tau_{353}$ across this region \citep{2014AeA...571A..11P}.

\subsection{Gamma-ray model}\label{sec:gmodel}

\begin{figure*}
  \centering 
  \includegraphics[width=\hsize]{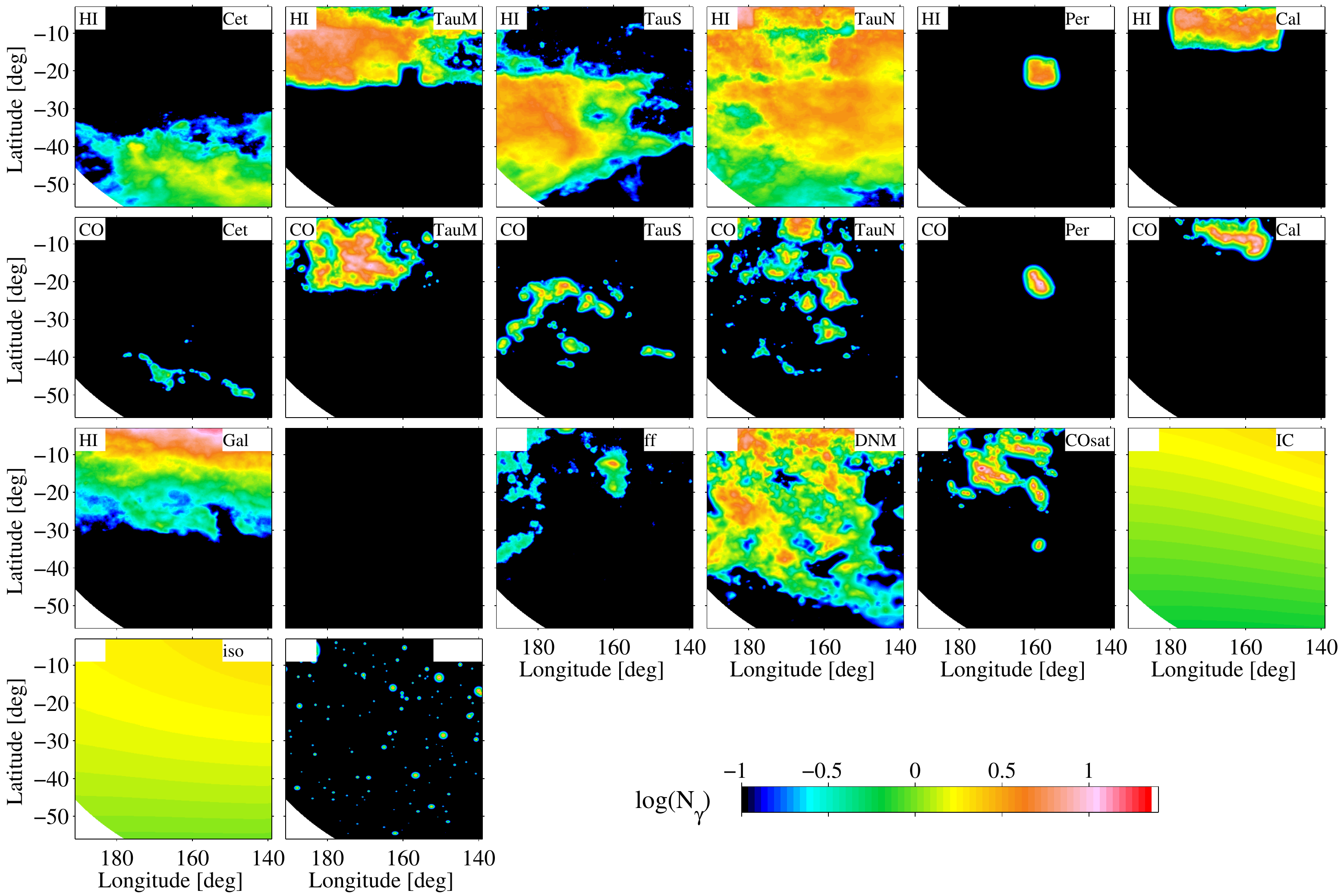}
  \caption{ Photon yields arising in the \g-ray model in the 0.4--100 GeV band from the \hi (HI label) and CO-bright (CO label) phases of the Cetus, Main Taurus, South Taurus, North Taurus, Perseus, California, and Galactic disc clouds, from the ionised gas (ff label), from the DNM and \cosat gas column densities, from the Galactic IC emission, and from the isotropic background (iso label) and \g-ray point sources.}
  \label{fig:gamMod}
\end{figure*}

Earlier studies have indicated that the bulk of the Galactic CRs radiating at 0.4--100 GeV have diffusion lengths far exceeding typical cloud dimensions and that they permeate all the \hi-bright, DNM, and CO-bright gas phases. The observed \g-ray emission can therefore be modelled, to first order, by a linear combination of the same gaseous components as in the dust model. We have assumed that the emissivity spectrum of the gas follows the average one obtained in the local ISM \citep[$q_{\rm LIS}(E)$,][]{2015ApJ...806..240C}, but we have left a free normalisation in each energy band to account for possible deviations in CR density and spectrum. The model includes other radiation components such as the Galactic IC radiation, $I_{\rm IC}(l,b,E)$, the isotropic intensity mentioned above, $I_{\rm iso}(E)$, and point sources with individual flux spectra $S_j(E)$. We have verified that the soft emission from the Earth limb is not detected in the present energy range for the choice of maximum zenith angle. The soft and transient emission from Sun and Moon is not expected to be detected as the number of \g-ray photons they emit over 6 years is negligible compared to those of the ISM components in the energy range studied. 
The \g-ray intensity $I(l,b,E)$, expressed in cm$^{-2}$ s$^{-1}$ sr$^{-1}$ MeV$^{-1}$, can thus be modelled as : 
\begin{eqnarray}
      I(l,b,E) &=& q_{\rm LIS}(E) \,\times \, \bigr[\, \sum_{i=1}^7 q_{\rm{HI},i}(E) \, N_{\rm{HI},i}(l,b) \nonumber\\
      &+&\, \sum_{i=1}^7 q_{\rm{CO},i}(E) \, W_{\rm{CO},i}(l,b) + q_{\rm{ff}}(E)I_{\rm{ff}}(l,b) \nonumber \\ 
      &+&\, q_{\rm{DNM}}(E) \, \tau_{353}^{\rm{DNM}}(l,b) + q_{\rm{COsat}}(E) \, \tau_{353}^{\rm{COsat}}(l,b)\,\,\bigr] \nonumber \\
      &+&\, q_{\rm{IC}}(E) \, I_{\rm IC}(l,b,E) ) + q_{\rm iso}(E) \, I_{\rm iso}(E) \nonumber\\
       &+&\, \sum_j q_{S_j}(E) \, S_j(E) \, \delta(l-l_j,b-b_j) \nonumber\\
        &+&q_{S{\rm ext}}(E) \, S_{\rm ext}(l,b,E), 
\label{eq:modGam}
\end{eqnarray} 
with the $\tau_{353}^{\rm{DNM}}$ and $\tau_{353}^{\rm{COsat}}$ maps extracted from the coupled dust and \g-ray analyses (see Sect. \ref{sec:iter}). 

The input \qlis spectrum was based on four years of LAT data and on the correlation between the \g radiation and the \nhi column densities derived from the LAB survey, for a spin temperature of 140 K, at latitudes between 7\degr~and 70\degr~\citep{2015ApJ...806..240C}. The $q_{\rm{HI},i}$ scale factors in the model can therefore compensate for differences in the \hi data (calibration, angular resolution, spin temperature) and potentially for cloud-to-cloud variations in CR flux. Such differences will affect the normalizations equally in all energy bands whereas a change in CR penetration in a specific cloud will show as an energy-dependent correction. For each cloud, the average \g-ray emissivity spectrum per H atom in the atomic phase is estimated from the product of the \qlis spectrum and the best-fit $q_{\rm{HI},i}$ normalization. This emissivity can be used to estimate the gas mass present in the other DNM, CO, and \cosat parts of the cloud if one assumes a uniform CR flux across the whole structure.   

The model includes 126 points sources from the 3FGL catalogue \citep{2015ApJS..218...23A} inside the analysis region. Their flux spectra, $S_{j}(E)$, have been computed with the spectral characteristics provided in the catalogue. Their $q_{S_j}(E)$ normalization in the model allows for possible changes due to the longer exposure and different photon reconstruction dataset used in the present analysis (six years of Pass 8 data instead of four years of Pass 7 reprocessed data for 3FGL) and due to the use of a different interstellar background for source detection in 3FGL. The sources have been fitted simultaneously. The sources present in the 4\degr-wide peripheral band around the analysis region have been merged into a single map, $S_{ext}(l,b,E)$, and its normalization, $q_{\rm{S_{ext}}}$, has been left free in each energy band. The IC emission has been derived from a GALPROP model \citep{2012ApJ...750....3A} and isotropic emission has been determined over the whole sky with a different interstellar background model \citep{2015ApJ...806..240C}, so we have left their scaling free in each energy band. 

In order to compare with the LAT photon data in the different energy bands, we have multiplied by the exposure and convolved by the LAT PSF each model component. Figure \ref{fig:gamMod} gives the photon yields in the total energy band obtained by the fitting (Sec. \ref{sec:results}) from those components. It shows that the emission originating from the gas dominates over other types of emission and that the LAT angular resolution allows the spatial separation of the various clouds, and of the different gas phases within the clouds. The $q$ model parameters have been fitted to the LAT photon data in each energy band using a binned maximum-likelihood with Poisson statistics.

\begin{figure*}
  \centering      
  \includegraphics[width=\hsize]{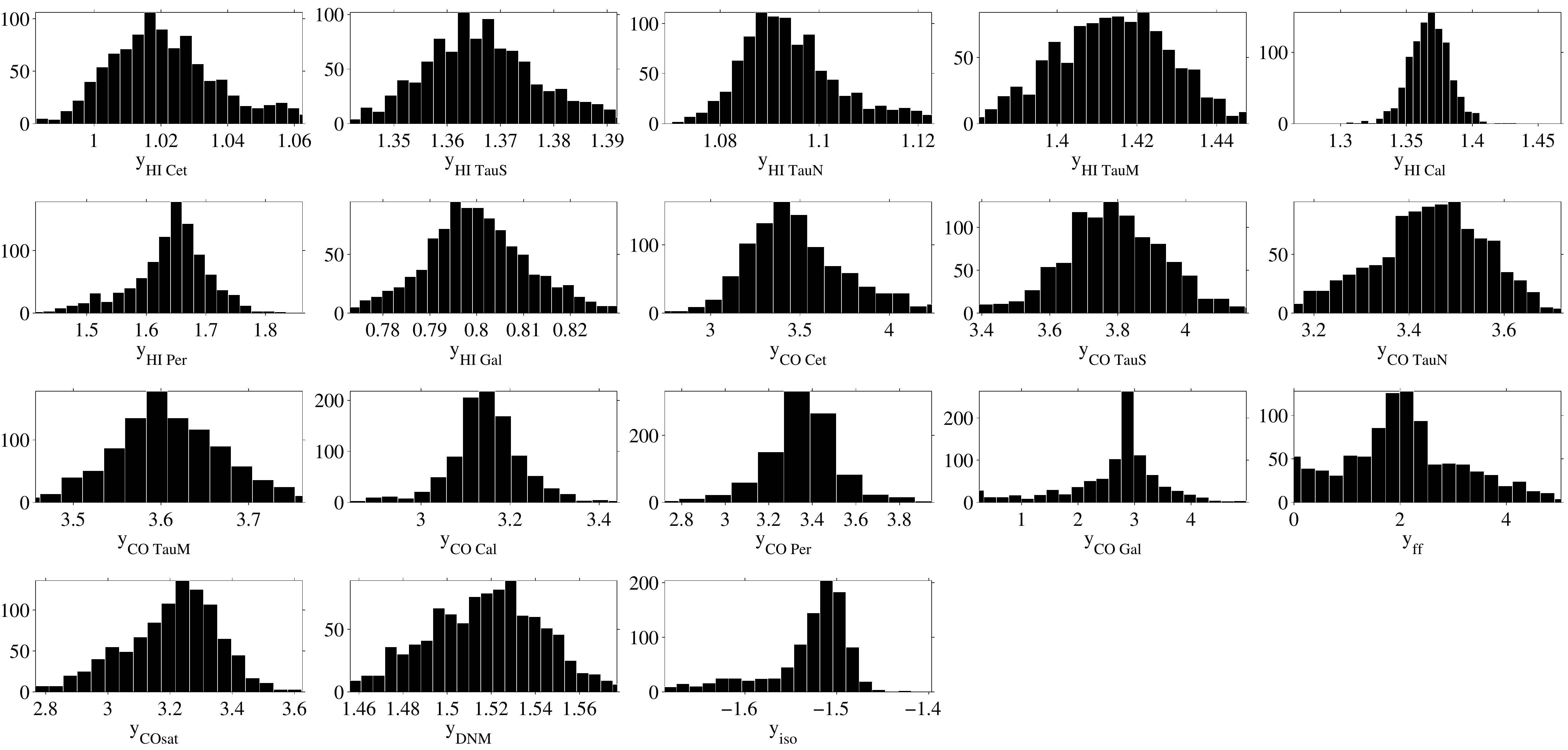}
  \caption{ Number distribution of the dust model coefficients over 1000 jackknife fits for an \hi spin temperature of 400~K. $y_{\rm{HI,i}}$, $y_{\rm{COsat}}$, and $y_{\rm{DNM}}$ are in units of $10^{-26}$ cm$^{2}$, $y_{\rm{CO,i}}$ in $10^{-6}$ K$^{-1}$ km$^{-1}$ s, $y_{\rm{ff}}$ in 3.8 $10^{-11}$ Jy$^{-1}$ sr  and $y_{iso}$ in $10^{-6}$.}
  \label{fig:yjack}
\end{figure*}
\begin{figure*}
  \centering                
  \includegraphics[width=\hsize]{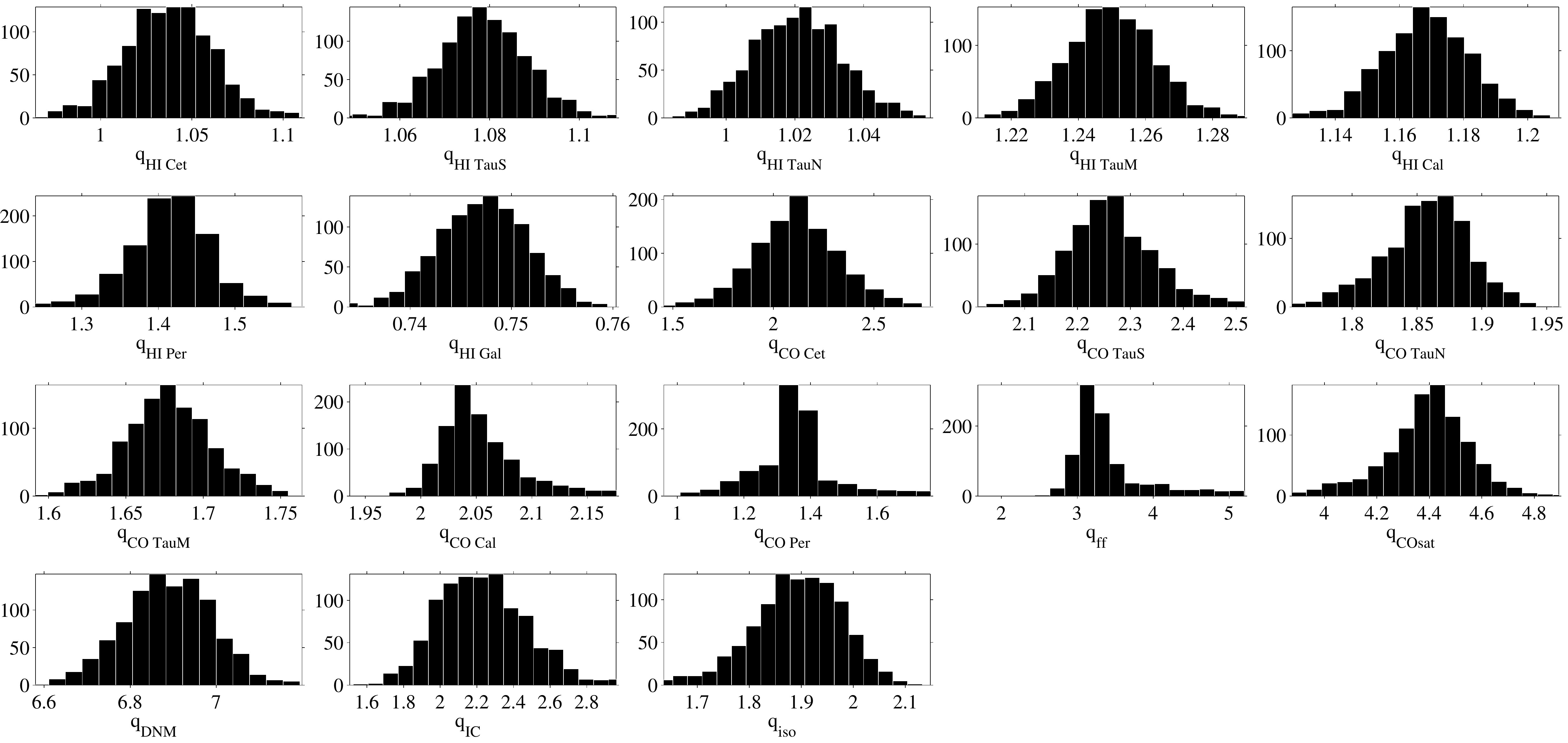}
  \caption{ Number distribution of the \g-ray model coefficients obtained for 1000 jackknife fits for an \hi spin temperature of 400~K. $q_{\rm{CO,i}}$ are in units of 10$^{20}$ cm$^{-2}$ K$^{-1}$ km$^{-1}$ s, $q_{\rm{ff}}$ in  3.8 10$^{15}$ cm$^{-2}$ Jy$^{-1}$ sr, $q_{\rm{DNM}}$ and $q_{\rm{COsat}}$ in $10^{25}$ cm$^{-2}$.  $q_{\rm{HI,i}}$, $q_{\rm{IC}}$ and $q_{iso}$ are normalisation factors.}  
  \label{fig:qjack}
\end{figure*}

\subsection{Analysis iterations}\label{sec:iter}

An important aspect of our analysis is the iterative coupling of the dust and \g-ray models in order to extract the common DNM and \cosat gas components which are present in both datasets, but for which we have no a priori templates. Both components show up as positive residuals over the expectations from the \hi-bright, CO-bright, and free-free-bright gas components. To extract them, we have built maps of the positive residuals between the data (dust or \g rays) and the best-fit contributions from the \nhi, \wco, free-free, and ancillary (other than gas) components. We have kept only the positive residuals above the noise (see below). We have separated the residuals into DNM and \cosat maps according to the \wco intensity in each direction (below and above $7$~K km/s for the DNM and \cosat, respectively). The DNM and \cosat templates estimated from the dust are provided to the \g-ray model ($\tau_{353}^{\rm{DNM}}$ and $\tau_{353}^{\rm{COsat}}$ in equation \ref{eq:modGam}); conversely, the DNM and \cosat column densities derived from the \g rays are provided to the dust model ($N_{\rm H}^{\rm{DNM}}$ and $N_{\rm H}^{\rm{COsat}}$ in equation \ref{eq:modDust}). We have started the iterative process by fitting the dust optical depth with the \hi, CO, free-free, and isotropic components to build the first DNM and \cosat maps for the \g-ray model. We have then iterated between the \g-ray and dust models until a saturation in the log-likelihood of the fit to \g-ray data is reached (from the third to the fourth iteration). 

The estimates of the $q$ and $y$ model coefficients and the DNM and \cosat templates improve at each iteration since there is less and less need for the other components, in particular the \hi and CO ones, to compensate for the missing gas. They still do at some level because the DNM and \cosat templates provided by the \g rays or dust emission have limitations (e.g., dust emissivity variations, limited \g-ray sensitivity). 

Care must be taken in the extraction of the positive residuals because of the Gaussian noise around zero. A simple cut at zero would induce an offset bias, so we have de-noised the residual maps using the multi-resolution support method implemented in the MR filter software \citep{1998AeAS..128..397S}. The stability of the iterative analysis and the results of the dust and $\gamma$-ray fits are discussed in the following sections.  

\section{Results}\label{sec:results}

\begin{figure*}[!t]
  \centering   
  \includegraphics[width=\hsize]{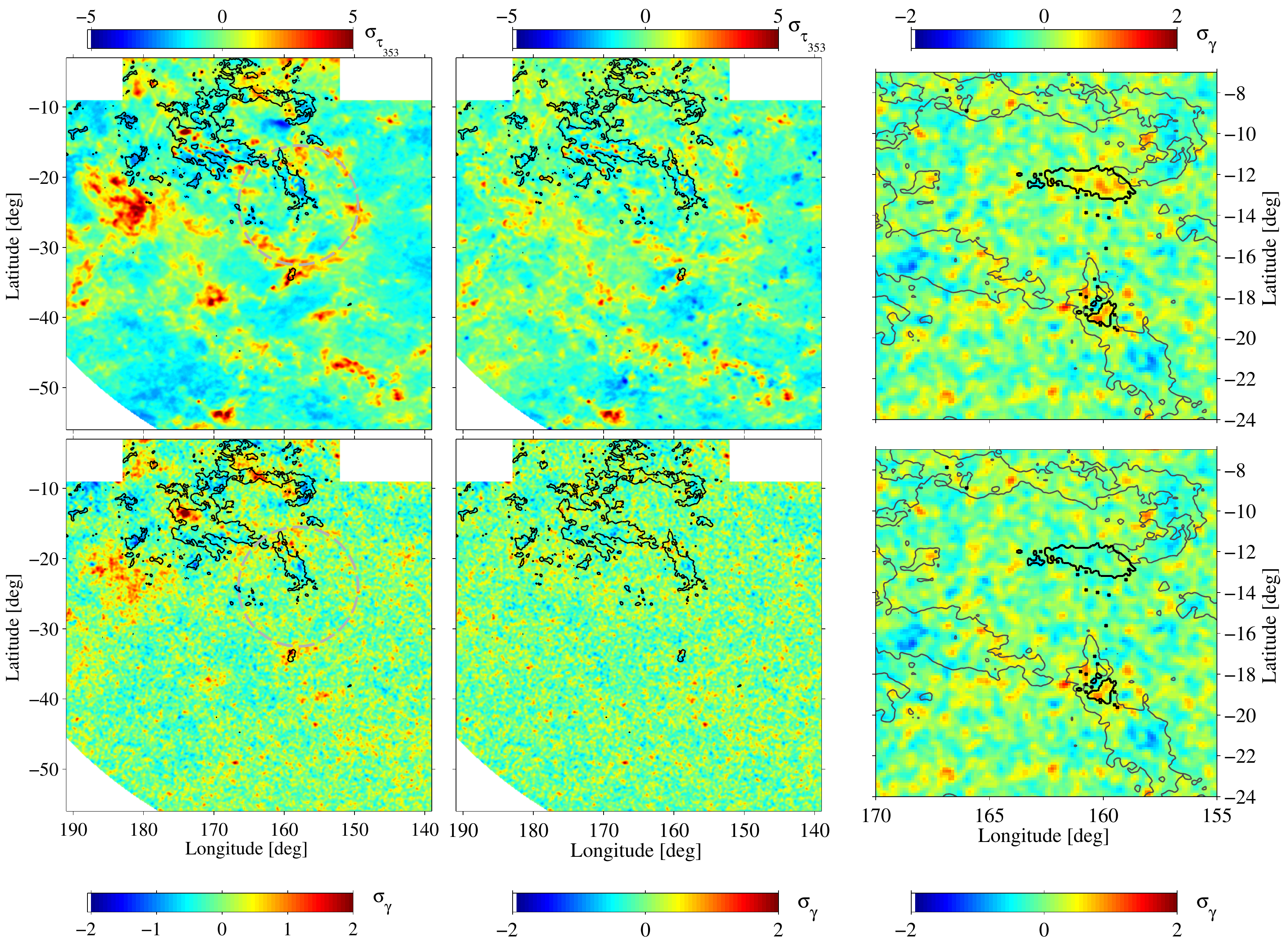}
  \caption{ Residual maps (data minus best-fit model,  in sigma units) in dust optical depth ($\sigma_{\tau_{353}}$) and in \g-ray counts in the 0.4--100 GeV band ($\sigma_{\gamma}$) obtained when including different sets of gaseous components in the models: without (\textit{left}) and with (\textit{middle}) the DNM and COsat components in addition to the \hi, CO, and free-free templates; a close-up view (\textit{right}) of the \g-ray residuals without (\textit{top}) and with (\textit{bottom}) the free-free map. The grey circles in the left-hand column highlight a DNM shell. The thin black contours outline the shape of the CO clouds at the 7 K km s$^{-1}$ level chosen to separate DNM and \cosat components. The thick black contours in the right-hand column outline the shape of the NGC 1499 and G159.6-18.5 \hii regions at the 2$\times 10^4$ Jy sr$^{-1}$ level in free-free emission at 70 GHz. The models are derived for an \hi spin temperature of 400~K.}  
  \label{fig:allRes}
\end{figure*}

\subsection{\hi optical depth correction}
The unknown level of \hi opacity in the different maps induces systematic uncertainties on the \nhi column densities, therefore on the \hi contributions to the models. The \g rays can help constrain the average level of \hi optical-depth correction applicable to the whole region by comparing the $T_{\rm{S}}$-dependent contrast of the \nhi maps with the structure of the  \g-ray flux emerging from the \hi gas. We have not tested different spin temperatures for each cloud complex. 
The maximum log-likelihood value of the \g-ray fit peaks for an \hi spin temperature near 400~K (see Fig. \ref{fig:dlnL_Tpsin} in Appendix \ref{sec:AnnexT}). In the following, we present the results obtained for this temperature, unless otherwise mentioned. 

Spin temperatures are generally measured from pairs of \hi emission and absorption spectra against bright radio sources \citep{2003ApJS..145..329H,2004JApA...25..143M,2011ApJ...737L..33K,2013MNRAS.436.2352R,2015ApJ...804...89M}. Line-of-sight harmonic means assume a single spin temperature along a sight line over the whole \hi line or per velocity channel, as in the present work (monophasic $T_{\rm{S}}$). The average value near 400 K that we find is consistent with the sparse distribution of previous monophasic spin measurements in this region \citep{2004JApA...25..143M,2011ApJ...737L..33K} and with the temperature span of 200--600~K found at high Galactic latitudes for the range of \nhi column densities dominating the present maps.
But the monophasic assumption is known to bias the spin temperatures to higher values and to characterise the mixture of the cold and warm neutral mediums (CNM and WNM) along the line of sight, rather than the physical temperatures of individual structures \citep{2015ApJ...804...89M}. Monophasic temperatures tend to decrease with increasing \nhi column densities because of the increasing proportion of CNM along the line of sight \citep{2011ApJ...737L..33K}.

\subsection{Best fits and jackknife tests}\label{sec:jack}

The best dust and \g-ray fits that could be achieved with the models described by Equations \ref{eq:modDust} and \ref{eq:modGam} include all the free-free, \hi, DNM, CO, and \cosat templates for the gaseous components. Table \ref{tab:fitscoef} gives the corresponding best-fit coefficients and Sec. \ref{sec:finalMod} discuss the goodness of fit with these models. Sections \ref{sec:dnmcosat} and \ref{sec:ffres} further discuss how the models have improved by adding gas templates other than \hi and CO. We focus here first on the determination of the uncertainties associated with each parameter in the models. 

The large spatial extents of the maps and the existence of tight correlations between the gas, dust, and \g-ray distributions yield small statistical errors on the best-fit coefficients. They have been inferred from the information matrix of the fit \citep{1985AeA...150..273S} and they include the effect of correlations between parameters. The gas parameters of the local clouds have precisions of 1--5\% and 1--4\% in dust and \g rays, respectively. Only the small contributions of the \hii regions and of the Cetus CO clouds to the \g-ray data have respective uncertainties of 10\% and 14\%. 

We have checked the magnitude of systematic uncertainties in our linear modelling approximations, hence of spatial changes in the model and/or in the mean level of \hi and CO self-absorption. We did so by repeating the last dust and \g-ray fits a thousand times over subsets of the analysis region, namely after masking out 20\% of the pixels with a sum of 2.625$^\circ$-wide, randomly selected squares. In \g rays, the jackknife tests have been performed only for the total 0.4--100 GeV energy band. 

Figures \ref{fig:yjack} and \ref{fig:qjack} show the distributions thus obtained for the best-fit coefficients that are significantly detected. Most of them show Gaussian-like distributions, thereby indicating that the results presented in Tables \ref{tab:fitscoef} are statistically stable and that the average coefficients that describe our models are not driven by subset regions in each cloud complex.
Several distributions exhibit long, non-Gaussian tails when the corresponding clouds subtend small solid angles (e.g., the \hii regions or the CO clouds from the Galactic disc or Perseus). The long tails reflect the indeterminacy of the parameter when a large fraction of their maps are masked. 

The standard deviations found in the jackknife distributions amount to 1--3\% and 1--4\% for the extended \hi and DNM components in dust and \g rays, respectively. The deviations are slightly larger, respectively 1--5\% and 2--10\%, for the more compact CO clouds in the local ISM. 


We have quadratically added the 1 $\sigma$ fitting error and the standard deviation of the jackknife distributions to give the statistical errors listed in Table C.1. The jackknife tests have been carried out for the total energy band. We have assumed the same relative deviations, $\sigma_q/q$, for the individual energy bands as for the total one since the jackknife samples test potential non-uniformities in the models at larger scales than the angular resolution of the data. 

The results indicate that the small contribution of the faint Cetus CO clouds to the total \g-ray photon counts is only detected below 4 GeV. Similarly, the faint CO clouds from the edge of the Galactic disc are not detected against the much larger \g-ray contribution from the Galactic atomic gas (see Fig. \ref{fig:gamMod}). In the dust fit, all components are significantly detected.

We have checked the convergence of the iterative scheme described in Sect. \ref{sec:iter} by monitoring the increase in the maximum log-likelihood value of the \g-ray fit as we progressed into the iterations until saturation. 
The log-likelihood ratio $2\Delta \ln(L)$ of 1180 between the last and first iterations indicates a large improvement in the quality of the fits because of a better separation of the different gas structures. The initial fits yielded too-large \hi and CO parameters to compensate for the missing gas. As the construction of the DNM and \cosat maps improved through the iterations, the bias on the \hi and CO parameters relaxed. Between the first and last iterations, the \hi and CO \g-ray emissivities dropped by 5-21\% and 4-22\%, respectively, depending on the cloud complex and its spatial overlap with DNM and \cosat structures. Similarly, the dust opacities of the \hi and CO clouds decreased by 10-42\% and 36-87\%, respectively. 

\begin{figure}
 \centering
 \includegraphics[width=\hsize]{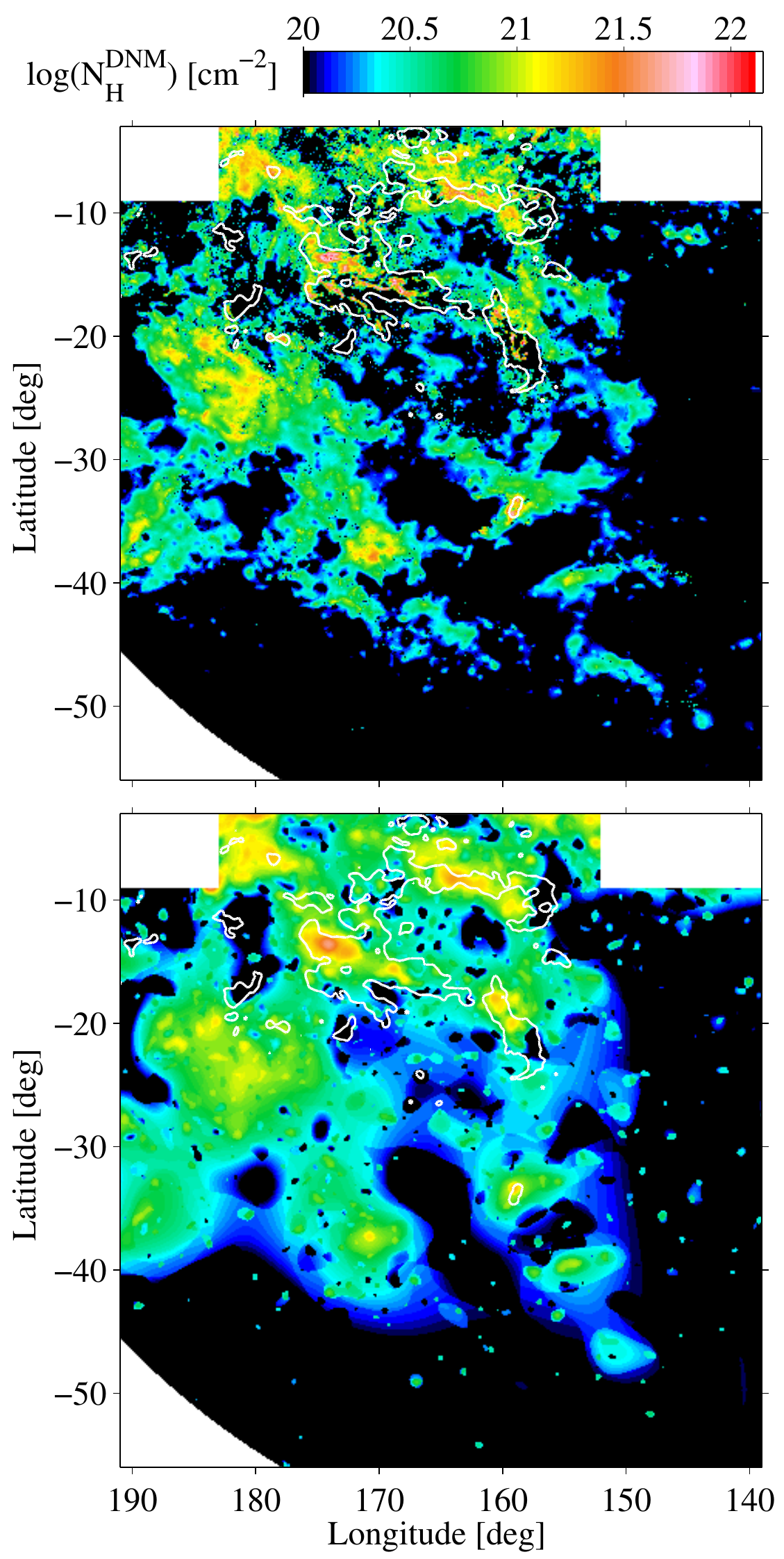}
 \caption{ Hydrogen column density maps in the DNM and \cosat components, derived from the dust (\textit{top}) and \g-ray (\textit{bottom}) data. The demarcation between the two components is outlined by the white contour at 7 K km/s in CO line intensity.}
  \label{fig:DNM}
\end{figure}

\subsection{Detection of DNM and \cosat gas}
\label{sec:dnmcosat}

Prior to building the complete models, we have checked for the presence of substantial amounts of gas that are not traced by the \hi and CO line intensities or free-free emission, but that are perceptible in parallel in the independent \g-ray and dust data sets. To do so, we have performed the dust and \g-ray fits with only the \hi, CO, and free-free templates for the gaseous components. We find that the data exceed the model predictions in several extended regions (see the positive residuals in the left column of Figure \ref{fig:allRes}) and that the excesses correlate in space in the independent dust and \g-ray data. Adding the DNM and \cosat templates to the models significantly improves the quality of the fits (see the middle column of Figure \ref{fig:allRes}). The log-likelihood ratio of 6427 between the \g-ray fits with and without the DNM and \cosat components implies that those components are detected at a formal confidence level greater than 80\,$\sigma$ \citep{1933RSPTA.231..289N}. Hadronic interactions between cosmic rays and dust grains, or IC up-scattering of the dust thermal radiation would yield too few \g rays to explain such a correlation \citep{2005Sci...307.1292G}. A coincident variation both in CR density and in dust-to-gas ratio is very improbable, so the CRs and dust jointly reveal gas in excess of the \hi, CO, and ionised expectations in those directions.
The spectrum of the \g-ray emission associated with both components is consistent with that found in the other gas phases over the 0.4-100 GeV energy range (see Sect. \ref{sec:qHI}). This gives further support to a gaseous origin of the \g-ray emission in the DNM and \cosat structures and to a smooth penetration of the CRs from the atomic envelopes to the dense parts of the clouds with saturated CO lines.

A log-likelihood ratio of 1404 between the models with different or with equal emissivities for the DNM and \cosat components strongly supports their separation into distinct components. The DNM map traces column densities of dense \hi and/or diffuse \hd at the atomic--molecular transition of the clouds whereas the \cosat map reveals dense \hd in excess of that linearly traced by \wco because of the saturation of the CO lines (\wco remains constant as \nhd keeps increasing). Without kinematic information, we cannot separate the DNM and \cosat contributions from the different cloud complexes when they overlap in direction. To extract the column densities displayed in Fig. \ref{fig:DNM}, we have assumed that the DNM and \cosat gas is pervaded by the average CR density measured in the atomic phase of the different clouds in the region \citep{2005Sci...307.1292G,2015AeA...582A..31A}. We show in Sect. \ref{sec:qHI} that the small cloud-to-cloud dispersion in \g-ray emissivity per gas nucleon justifies the use of the average. 
We defer the discussion of the DNM and \cosat gas masses and their relation to the \hi-bright, $^{12}$CO-bright, and $^{13}$CO-bright phases to a companion paper. 

\subsection{Detection of ionised gas}
\label{sec:ffres}

We have also checked the significance of the addition of the free-free template to the final \g-ray and dust models, and we have tested the gaseous origin of the corresponding \g-ray signal. We develop both points in this section.

The log-likelihood ratio of 116 between the models with and without the free-free template supports a \g-ray detection of the \hii regions at a formal confidence level greater than 10$\,\sigma$. The close-up view in Fig. \ref{fig:allRes} shows the clear detection of a \g-ray excess toward the bright NGC 1499 region and the more marginal gain in adding the small column densities of ionised hydrogen from the fainter G159.6-18.5 region. By replacing the free-free map by a broad Gaussian source located toward the centre of NGC 1499, we have checked that the fit significantly requires an extended excess (7\,$\sigma$), with a FWHM comprised between $1.4^{\circ}$ and $3.6^{\circ}$ at the 95\% confidence level. The Gaussian source, however, yields a poorer fit than the more elongated free-free emission. 

The two \hii regions, NGC 1499 and G159.6-18.5, are respectively excited by a giant and a main-sequence O star. Part of their intense stellar light is upscattered to \g rays by the local CR electrons \citep{2007Ap&SS.309..359O}. Given the distance, luminosity, and effective temperature of each star, we have calculated the IC \g-ray flux produced in its vicinity and we have integrated the flux over the angular sizes of the \hii region, in the 0.4--100 GeV energy band. We find IC fluxes of ${\sim}2.9\times 10^{-11}$~cm$^{-2}$~s$^{-1}$ and ${\sim}5.9\times 10^{-12}$~cm$^{-2}$~s$^{-1}$, respectively, that cannot account for the \g-ray fluxes of $(2.5 \pm 0.4)\times 10^{-8}$~cm$^{-2}$~s$^{-1}$ and $(2.3 \pm 0.4)\times 10^{-9}$~cm$^{-2}$~s$^{-1}$ seen in correlation with the free-free emission. 
Hence, the \g-ray excesses seen toward NGC 1499 and partially toward G159.6-18.5 are more likely due to hadronic interactions of the local CRs in the ionised gas.

The weak correlation between the dust optical depth \taunu and the free-free template highlighted by the jackknife test probably reflects the difficulty of estimating \taunu toward regions of strong spectral gradients (see Fig.~\ref{fig:PLK4}). In particular because the $\beta$ index and the grain temperature have been derived at different angular resolutions \citep{2014AeA...571A..11P}.

Assuming a uniform CR flux in the \hi and \hii gas phases, we have used the average \g-ray emissivity of the atomic gas in the region (see Sect. \ref{sec:qHI}) and the best-fit $q_{\rm ff}$ parameter obtained for the 0.4--100 GeV band to convert the free-free intensities to hydrogen column densities and compare them with the values expected from equation \ref{eq:NHII}. The electron temperature of \hii regions is known to vary with Galactocentric radius because of the metallicity gradient \citep{2012MNRAS.422.2429A}. Adopting a mean value in the solar neighbourhood between 7000~K and 8000~K, we find an average electron density of ($4.3 \pm 0.6)$~cm$^{-3}$ in the \hii regions sampled here. 

At the distance $d\approx 380$ pc to $\xi$ Persei inside NGC 1499 \citep{2007AeA...474..653V}, the 80\arcmin\, radius of the ionised region is $L=9$ pc. The fractional thickness of the shell, relative to $L$, is estimated to be $l=0.38$ \citep{1976PASJ...28..437S} and the nebula subtends a solid angle $4 \pi \sigma$ with $\tilde{\sigma}=0.21$  at the exciting star. The H$_\alpha$ flux is corrected for the interstellar absorption as $F_{\alpha 0}= 10^{0.292 A_V}F_{\alpha}$, with $A_V$ the visual extinction up to 500 pc derived from the 3D extinction map of \cite{2015ApJ...810...25G}.
According to \cite{1976PASJ...28..437S} for equal electron and ion volume densities, the mean electron density of a shell-like nebula such as NGC 1499 can be expressed as $\overline{n}_e=SV^\frac{1}{2}$ with the effective volume
\begin{equation}
 V=\frac{4}{3}\pi L^3 \tilde{\sigma} [1-(1-l)^3],
\end{equation}
and the excitation parameter
\begin{equation}
\frac{S}{{\rm cm}^{-3/2} }= \left( \frac{1}{9.46 \times 10^{-60}}\right) \left( \frac{F_{\alpha 0}}{1\,{\rm erg\,cm}^{-2}\,{\rm s}^{-1}}\right) 
\left( \frac{d}{1\,{\rm pc}}\right)^2 \left( \frac{T_e}{1\,{\rm K}}\right)^{0.87}. 
\end{equation}

The electron temperature $T_e$ of the nebula can be estimated by comparing the free-free and H$_\alpha$ emissions. The H$_\alpha$ intensity in Rayleigh units is expressed as a function of the emission measure $\int n_e n_i ds$ as :
\begin{equation}
I_{H_{\alpha}}= \frac{1}{2.75} \left( \frac{T_e}{10^4\,{\rm K}} \right)^{-0.9} \int n_e n_i ds. 
\label{eq:IHa}
\end{equation}
Combining equations \eqref{eq:Iff} and \eqref{eq:IHa}, we obtain 
\begin{equation}
T_e= 2.43 \,{\rm K} \left( \frac{I_{\rm{ff}}}{I_{H_{\alpha}}} \right)^{\frac{1}{0.62}}.
\label{eq:Te}
\end{equation}
Given this temperature, we can estimate the mean electron density.

If we apply those calculations to NGC 1499 within a low contour at $I_{\rm{ff}}=2\times10^4$ Jy sr$^{-1}$, we find a mean temperature of $(6100 \pm 2700)$ K and a mean electron density $\overline{n}_e=(15.1\pm 6.7) $ cm$^{-3}$. If we adopt a more restrictive contour where the \g-ray residuals exceed 0.8$\sigma$ in the best-fit model without the free-free template, we find $\overline{T}_e=(7700 \pm 1700)$ K and a mean density $\overline{n}_e=(5.0\pm 1.1) $ cm$^{-3}$ in good agreement with the gas density estimate inferred from the correlation between the \g rays and the nebular free-free emission. This agreement gives further weight to a hadronic origin of the \g rays associated with NGC 1499.

\subsection{Final model residuals}\label{sec:finalMod}

\begin{figure*}[!t]
  \centering     
  \includegraphics[scale=0.6]{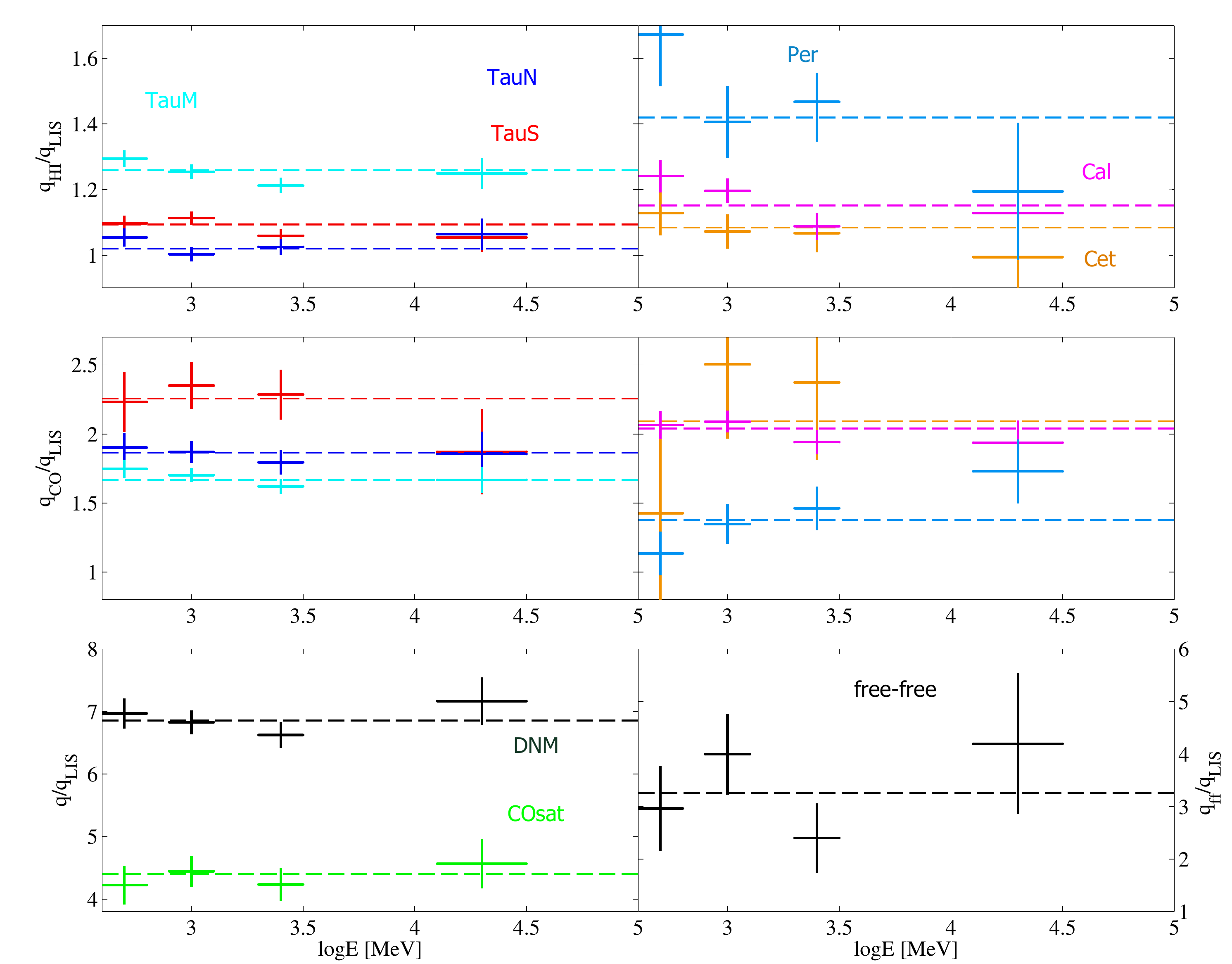}
  \caption{ Spectral evolution, relative to the local interstellar spectrum \qlis, of the \g-ray emissivities of the local gas components. \qco is given in units of $10^{20}$cm$^{-2}$ (K km s$^{-1}$)$^{-1}$ and \qdnm in $10^{25}$cm$^{-2}$. The \qhi normalizations are given for a spin temperature of 400~K. The dashed lines compare the results obtained for the whole 0.4--100 GeV band to the results of the four independent bands. All error bars are symmetrical; horizontal error bars at the boundaries are truncated.}   
  \label{fig:qfitTau}
\end{figure*}

The residuals obtained between the dust and \g-ray data and the best fits that include all gaseous components are presented in the middle column of Fig. \ref{fig:allRes}. They show that the linear model provides an excellent fit to the \g-ray data in the overall energy band, as well as in the separate energy bands which are not shown. The residuals are consistent with noise at all angular scales except, marginally, toward the brightest CO peaks where the model tends to over-predict the data. Significant positive residuals remain at small angular scales in the dust fit. They closely follow the spatial distribution of the DNM and they are likely due to the limitations in angular resolution and in sensitivity of the \g-ray DNM template compared to its dust homologue. Small-scale clumps in the residual structure can also reflect localised variations in dust properties per gas nucleon that are not accounted for in the linear models, in particular toward the brightest CO regions. These effects are discussed in Sect.~\ref{sec:dust}.

A wide, 17\degr-diameter shell, centred on $l=158^{\circ}$ and $b=-24^{\circ}$ is apparent in the initial dust residuals, and also marginally in the \g-ray residuals (see the left column of Fig. \ref{fig:allRes}). The shell is rich in DNM gas (see Fig. \ref{fig:DNM}). We have found no coincident structure at other wavelengths that could explain this shell in terms of a nearby \hii region or old supernova remnant. 

\section{$\gamma$-ray emissivity of the gas}
\label{sec:qHI}

The \g-ray emissivity spectra of the local gaseous components are presented in Fig. \ref{fig:qfitTau}. We note that all the \qhi normalizations exceed one because the present fits preferred a larger \hi spin temperature than the value of 140~K adopted for the derivation of \qlis across the whole sky off the Galactic plane \citep{2015ApJ...806..240C}. 
We find no significant spectral variations except in Perseus for which we observe opposite trends for \qhi  decreasing and \qco increasing with energy. This behaviour can be attributed to the cross talk between the very compact set of \hi and CO clouds in Perseus as the LAT PSF degrades at low energies.

The results indicate that the CR population permeating the various phases of the different clouds has the same energy distribution as the average in the local ISM \citep{2015ApJ...806..240C}. 
We find no spectral signature that would betray exclusion or concentration processes in a particular cloud or with increasing gas density, from the diffuse atomic gas in Cetus up to the densities of $10^{3-4}$~cm$^{-3}$ sampled by 2.6-mm CO line emission. Together with similar findings in the Chamaeleon clouds \citep{2015AeA...582A..31A}, these results support the theoretical predictions that CRs with energies between a few hundreds of MeV and a few tens of GeV smoothly diffuse through the \hi-bright and CO-bright parts of the ISM \citep{1976AeA....53..253S,1978AeA....70..367C,2011AeA...530A.109P}, up to column densities of $10^{21-22}$ cm$^{-2}$ \citep{2015MNRAS.451L.100M}. The volume densities of the gas covered by \hi and CO observations are modest, but they hold the bulk of the interstellar gas mass, \g rays should reliably trace most of the interstellar gas.

\begin{figure}
  \centering     
  \includegraphics[width=\hsize]{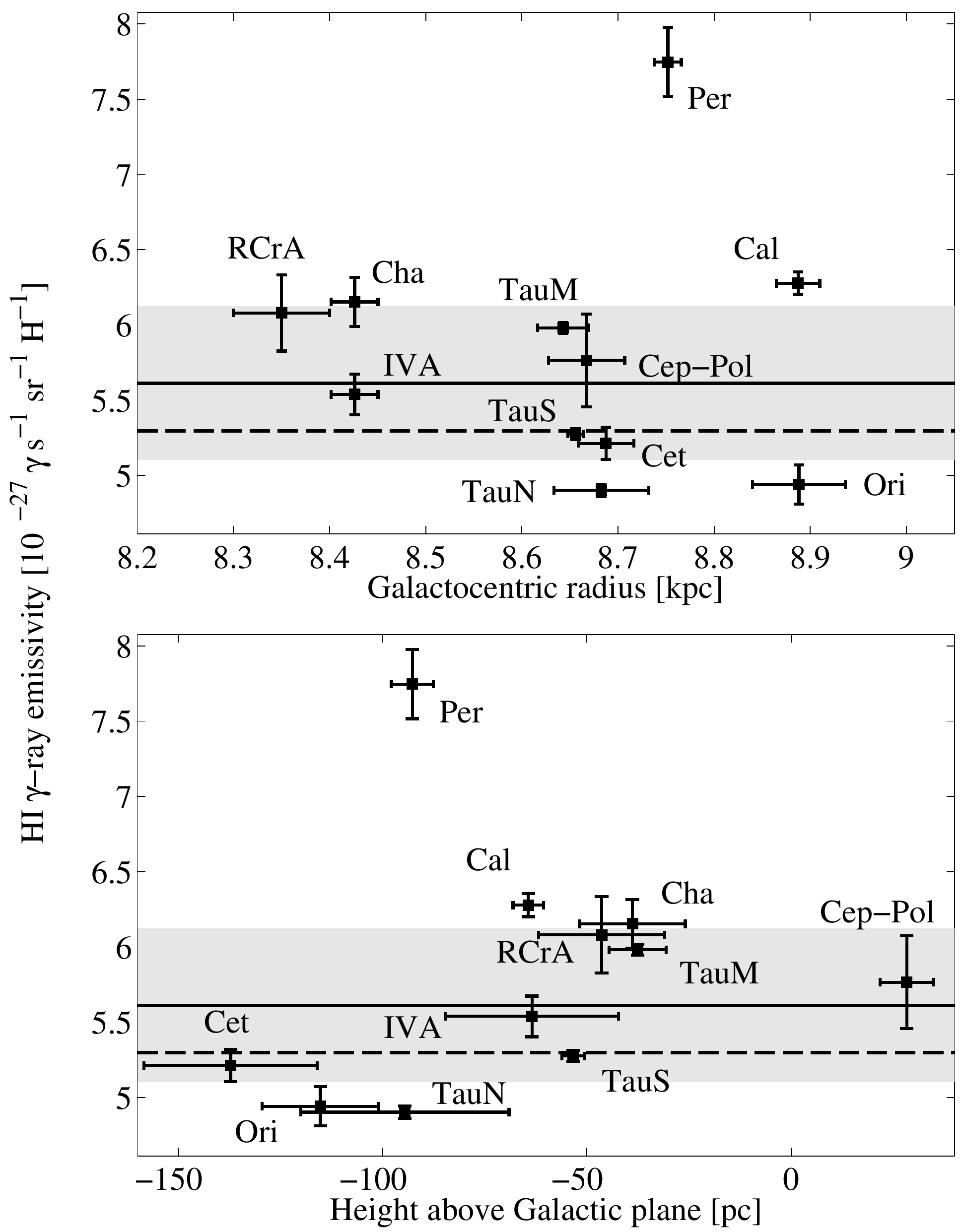}
  \caption{ Distribution with Galactocentric radius (top) and altitude above the Galactic disk (bottom) of the 0.4--10 GeV emissivities measured in the atomic gas of nearby clouds for \hi spin temperatures between 125 and 150~K. The solid and grey band respectively give the average emissivity and ${\pm}$\,1 rms dispersion in the sample. The dashed line marks the average emissivity measured across the sky at $|b| \ge 7^{\circ}$ \citep{2015ApJ...806..240C}.  }  
  \label{fig:Rgal}
\end{figure}

We can compare the \g-ray emissivities measured in the anticentre clouds for a spin temperature of 150~K with the average value found in the local ISM for  $T_{\rm spin} = 140$~K \citep{2015ApJ...806..240C}, as shown in Fig. \ref{fig:Rgal}.  
The 35$\pm 1 \%$ lower emissivity that we find for the atomic gas in the Galactic-disc background is not useful for comparison as it integrates all distances beyond the local ISM, through regions of decreasing CR density along lines of sight outward through the Galaxy \citep{2010ApJ...710..133A,2011ApJ...726...81A} and to high altitudes above the Galactic disc \citep{2015ApJ...807..161T}.
Within a few hundred parsecs, we find small cloud-to-cloud differences in \g-ray emissivity. The Cetus, South Taurus, and North Taurus clouds compare well with the average in the local ISM whereas the California, Main Taurus and Perseus clouds appear to be $15\,{\pm}\,3$\%, $12\,{\pm}\,2$\%, and $36\,{\pm}\,6$\% more emissive than the average, respectively. 
Except for Perseus, these variations are commensurate with previous measurements. Figure \ref{fig:Rgal} shows a cloud-to-cloud dispersion of $\pm$9\% that does not relate to the cloud altitude above the Galactic disc, nor to the cloud location with respect to the local spiral arm (Galactocentric distance). This dispersion largely stems from uncertainties in the derivation of the \nhi column densities. We note indeed that the \qhi emissivity is 10\% to 20\% larger in the Cepheus-Polaris, Main Taurus, California, RCrA and Chamaeleon clouds where \nhi column densities in the 80th percentile exceed $10^{21}$~cm$^{-2}$, a level at which \hi optical depth corrections become significant.

The emissivity in Perseus is at variance with the other nearby clouds, but less so at high energy where the LAT PSF allows a firmer separation between the compact atomic and molecular phases of the cloud. The origin of the high emissivity will be investigated using photon selections with improved angular resolution as soon as the photon statistics permit the analysis. For the anticentre region under study, we derive very consistent averages regardless of whether we include  the Perseus data to the other five measurements or not. We have used this average emission rate in the 0.4--100 GeV band, $\overline{q}_{\rm HI} = (6.0 \pm 0.3)\times 10^{-27} \gamma$~s$^{-1}$ sr$^{-1}$ H$^{-1}$, in the atomic gas to infer column densities in the other gas phases.

\section{X$_{\rm{CO}}$ factors} 

\begin{figure}
  \centering      
  \hspace{0.5cm}          
  \includegraphics[scale=0.32]{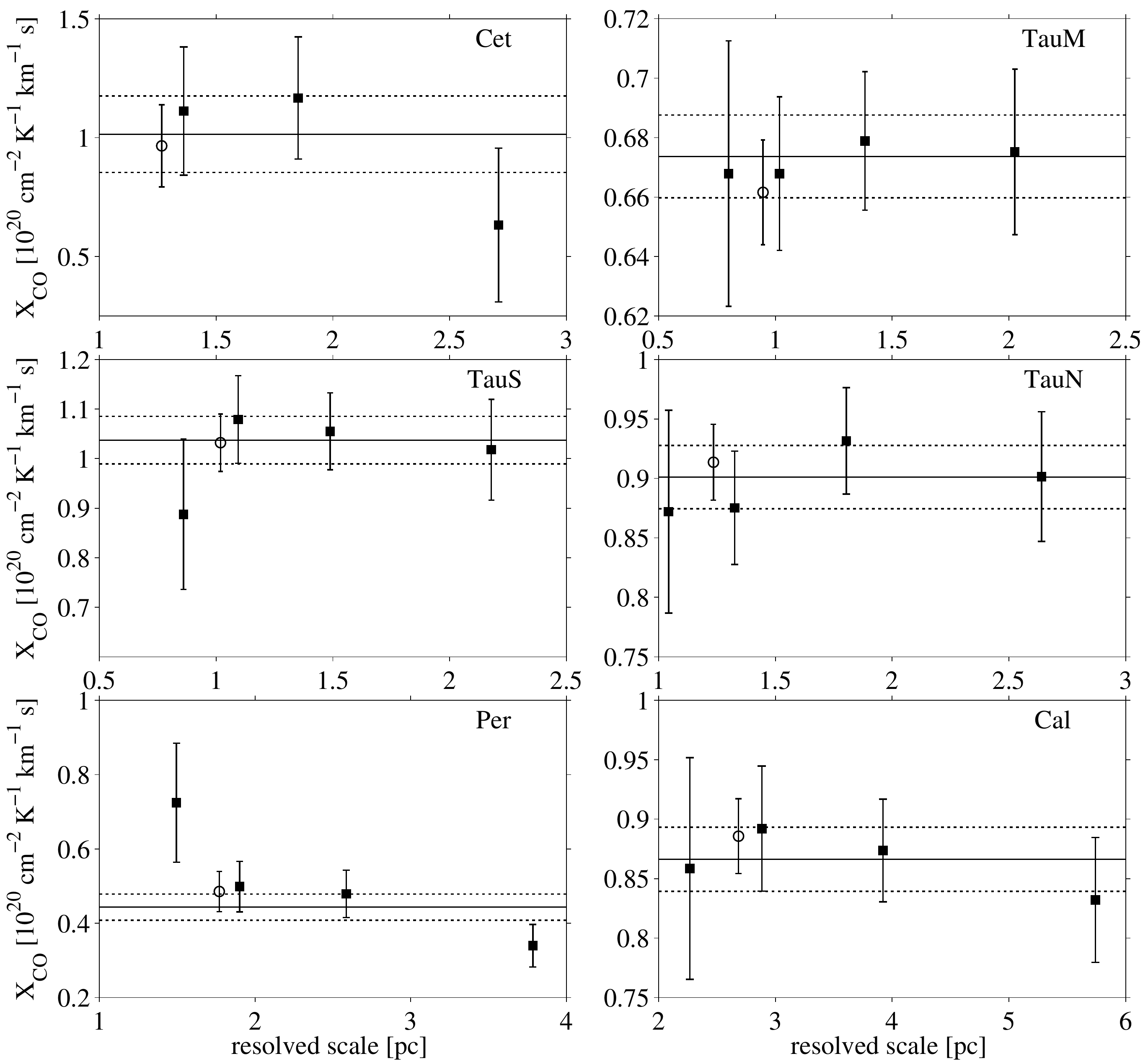}
  \hspace{0.5cm}
  \caption{ Evolution of the \xco factor, as measured in \g rays for various linear resolutions in the different clouds. The open circle marks the value obtained in the overall 0.4--100 GeV band, in close agreement with the weighted average of the four independent energy bands (black line) and its 1$\sigma$ errors (dashed lines).}
   \label{fig:res_scale}
\end{figure}

The \xco factor is often applied to convert \wco intensities to hydrogen-equivalent \nhd column densities in the molecular phase. In \g rays, under the assumption of a uniform CR flux in the \hi-bright and CO-bright phases, it is derived as $X_{\rm{CO}\gamma}= q_{\rm{CO}}/(2q_{\rm{H_{I}}})$, independently in each energy band. This ratio also assumes that helium is uniformly mixed by number with hydrogen in the ISM. Likewise, assuming a uniform dust-to-gas mass ratio and a uniform emission coefficient $\kappa_{353}$ of the grains at 353 GHz, the \xco factor is derived in the dust fit as $X_{\rm{CO}\tau}= y_{\rm{CO}}/(2y_{\rm{H_{I}}})$.
The results of the two methods for the six CO clouds selected in our analysis are given in Table \ref{tab:XCOfactors}. The \g-ray results in the individual energy bands are displayed in Fig. \ref{fig:res_scale}.

\input{XCO_IG.txt}

\subsection{X$_{\rm{CO}}$ measurements in \g rays} 

We find \xco values for the present clouds that compare well with previous \g-ray measurements in other nearby clouds. For \xco factors obtained with \textit{Fermi} LAT at comparable angular resolution, we can cite the values of $0.63 \pm 0.02^{+0.09}_{-0.07}$ in the Cepheus-Polaris complex \citep{2010ApJ...710..133A}, of $0.99 \pm 0.08^{+0.18}_{-0.10}$ in the RCrA cloud \citep{2012ApJ...755...22A}, of $1.21 \pm 0.02$ in the Orion clouds \citep{2012ApJ...756....4A}, and of $0.69\pm0.02^{+0.03}_{-0}$ in the Chameleon clouds \citep{2015AeA...582A..31A}, in units of \xcounit. 
We note that the \xco factors in those well-resolved clouds are all close to or lower than \xcounit. Table \ref{tab:Xco_history} lists historical \g-ray estimates of \xco. It shows that the estimates in nearby clouds have remained stable or only slightly decreased as the angular resolutions of the \hi, CO, and \g-ray surveys have improved. The lower level of cross-correlation between the \hi and CO structures on the one hand, a better separation of the \g radiation produced in the atomic and molecular phases on the other hand, and a lower contamination from unresolved \g-ray point sources has primarily improved the precision of the measurements. The latest analyses with $<$10.8\arcmin\, FWHM resolution in \hi, 8.5\arcmin\, resolution in CO, and \textit{Fermi} LAT data consistently yield low \xco values, ranging between 0.5 $\times$\xcounit and 1.2 $\times$\xcounit. Systematic uncertainties on the \nhi column densities due to \hi optical depth corrections impact the estimation of the \g-ray emissivities per gas nucleon. Analyses for different \hi spin temperatures show that the resulting systematic uncertainty on the \xco factors amount to (0.1-0.2) $\times$\xcounit.

The values listed in Table \ref{tab:Xco_history} highlight a recurrent discrepancy between the \xco measurements at small scale in nearby clouds and the two-to-three times larger values found at kiloparsec scales when averaging over a population of molecular complexes in spiral arms or in Galactocentric rings. We return to this point at the end of this section.

In Fig. \ref{fig:res_scale} we take advantage of the energy-dependent PSF of the LAT to probe various linear scales within the present nearby clouds.
The scale is derived using the half-width at the half-maximum of the PSF integrated over the energy band for the \qlis spectral shape. We find no evidence of \xco changes at parsec scales except in Perseus. 

Because the \qhi and \qco variations in Perseus are inversely coupled in energy, the resulting change in \xco is likely due to an increased level of cross-correlation between the compact \hi and CO phases as the LAT PSF degrades (Figs. \ref{fig:gamMod} and \ref{fig:qfitTau}). 
If we force the \g-ray emissivity of the \hi gas to be the same as the average found among the other clouds, we obtain a larger \xco value of $(0.68\pm0.04)\times$\xcounit, close to that measured above 4 GeV with the best LAT angular resolution. However, the use of this value implies a significant (4.6$\,\sigma$) degradation of the fit quality at lower energies. The likelihood analysis significantly supports a lower \xco factor in Perseus. A low value is also indicated by the average of $0.3\times$\xcounit found at 0.4-pc resolution in the dust, \hi, and CO study of \citet{2014ApJ...784...80L}.

Several observations and numerical models have suggested that the \xco factor changes with the density, turbulent velocity, and metallicity of the gas, and with the penetration of the CO-dissociating radiation inside a cloud \citep{1988ApJ...332..432P,2006MNRAS.371.1865B,2008ApJ...687.1075S,2010AeA...518A..45L,2011MNRAS.412..337G,2011MNRAS.412.1686S,2012AeA...541A..58L,2014ApJ...784...80L,2016MNRAS.455.3763B}. One therefore expects intrinsic \xco gradients in clouds. Calculations by \cite{2006MNRAS.371.1865B} show that \xco should dramatically drop inward from the outskirts of a molecular cloud as the photo-dissociating radiation gets absorbed. The \xco factor drops down to a minimum value at \av around 2 mag. The \xco ratio then rises again in the denser cores because of the saturation of the optically thick CO lines. One goal of our analysis is to explore whether these intrinsic variations can change the mean \xco value averaged over a cloud, depending on how diffuse it is. We cannot yet measure the steepest \xco gradients at the periphery of clouds (\av $\ll 1$ mag) near the sensitivity threshold of CO surveys because of the complex 3D interface with the abundant atomic phase. We cannot map \xco gradients pixel by pixel in \g rays because of the photon statistics, but our sample includes more or less translucent clouds of the local ISM and we can explore changes in the mean value of \xco for clouds in different states.

\begin{figure*}[!t]
  \includegraphics[scale=0.75]{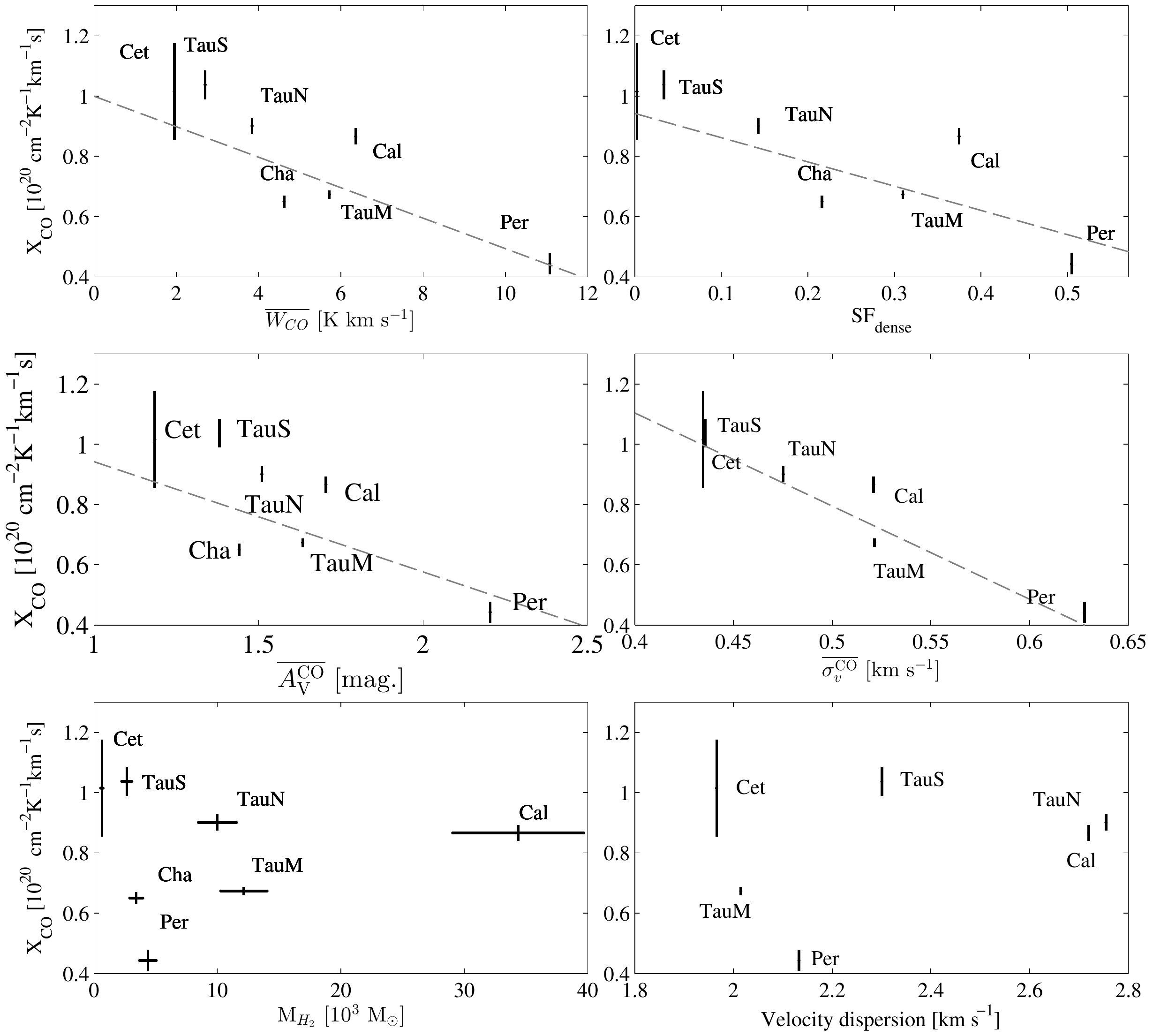}
  \caption{  Evolution of the X$_{\rm{CO}}$ factor measured in $\gamma$ rays as a function of the average  W$_{\rm{CO}}$ intensity, $\overline{W_{\rm CO}}$, the surface fraction of dense gas, $\rm{SF}_{dense}$, the mean visual extinction in the CO-bright phase, $\overline{A^{\rm{CO}}_V}$, the average CO line width, $\overline{\sigma^{\rm CO}_v}$,  the \hd mass in the CO-bright phase, $M_{{\rm H}_2}$, and the velocity dispersion of the CO line. Dashed lines give the best ($\chi^2$) linear regressions}
   \label{fig:XCOfactors}
\end{figure*}

Figure \ref{fig:XCOfactors} compares \g-ray measurements of \xco in nearby clouds from this work and \citet{2015AeA...582A..31A}. We restrict the sample to estimates obtained in the Chamaeleon and anti-centre clouds because they are based on the same analysis method, the same \g-ray energy bands to ensure the same angular resolution of the LAT, and the same sampling resolution in the \hi and CO data. They both include the DNM gas in the models. These clouds subtend large solid angles in the sky and their column density structures are well resolved. 

The distributions in Fig. \ref{fig:XCOfactors} show that the average \xco factor per cloud does not depend on the \hd mass mapped in CO, nor on the overall dynamics of the cloud characterised by the velocity dispersion of the CO lines. But \xco appears to depend on the cloud structure and in particular on its diffuseness. To reflect changes in the latter, we have explored several characteristics:
\begin{itemize}
\item the mean \wco intensity, \wcoavg, in a cloud, taken above 1 K km/s to avoid noise fluctuations; 
\item the surface fraction of dense regions with large \wco intensity within a cloud, $\rm{SF}_{dense}=S_{\rm{W_{CO}}>7 \; K km/s}/S_{\rm{W_{CO}}>1 \; K km/s}$ (with $S$ a solid angle), which gauges the relative weight of diffuse and dense molecular regions in the determination of \xco;
\item the mean visual extinction toward a cloud, \avavgco, taken from the NICER M2a 12\arcmin-resolution A$_{\rm J}$ map \citep{2016AeA...585A..38J}, translated into \av using a colour ratio of 3.55 according to the extinction law of \citet{1989ApJ...345..245C}. Only values in directions with \wco $>1$~K km/s and  \av $>0.8$~mag. have been retained in the average to avoid noise contamination;
\item the average CO line width, \sigvavg, in a cloud since more CO line photons can escape the cloud and ultimately be detected as the line width grows. The resulting increase in \wco intensity implies a decrease in \xco with \sigvavg \citep{2011MNRAS.415.3253S}.
\end{itemize} 

Figure \ref{fig:XCOfactors} shows that, for all these diagnostics, \xco tends to decrease from diffuse to more compact clouds.  The change in \xco is approximately a factor of two. The decrease remains significant if we use the Perseus \xco value found when forcing the \hi \g-ray emissivity to the local average, or if we exclude the Perseus data point. A uniform \xco factor is strongly rejected ($>30\sigma$)  for all the diffuse diagnostics, whether we modify or exclude the Perseus data point or not.

The trend is consistent with the expectation that \xco decreases from the diffuse envelopes, exposed to the photodissociating radiation field, to the shielded cores of the CO clouds \citep{2006MNRAS.371.1865B,2011MNRAS.412..337G,2011MNRAS.412.1686S,2016MNRAS.455.3763B}. Hence, the average \xco factor per cloud in our sample is seen to vary with the relative solid angle subtended by the diffuse and dense parts of the clouds, reflected in $\rm{SF}_{dense}$ and $\overline{W_{\rm CO}}$.
The turbulence level and the degree of virialization of the cloud (e.g., the ratio of kinetic to gravitational energy) also influences the average \xco factor by allowing the ISRF to penetrate more or less deeply into the turbulent layers of the cloud \citep{2006MNRAS.371.1865B,2011MNRAS.415.3253S,2016MNRAS.455.3763B}. The mean \xco factor per cloud in Fig. \ref{fig:XCOfactors} does decrease with the average line width of the CO lines. The present measurements also give weight to the idea that the mass or size of a cloud does not play an important role in determining the \xco factor \citep{2011MNRAS.415.3253S}.

According to \xco calculations as a function of visual extinction \av \citep{2006MNRAS.371.1865B}, \xco should drop by more than two orders of magnitude with increasing \av, down to a minimum at $1 \lesssim A_{\rm V} \lesssim 3$ mag. For long-lived clouds, exposed to the local cosmic-ray ionization rate of $1.4 \times 10^{-17}$ s$^{-1}$ \citep{2015ARAeA..53..199G} and to the local ISRF, the minimum \xco value reaches below \xcounit for hydrogen densities larger than $10^3$~cm$^{-3}$ and/or for turbulent velocities larger than 1 km/s. Yet, the measured CO lines have widths below 0.7 km/s and the clouds in our sample are not massive enough to sustain such large average densities through most of their volume. Moreover, models of dense clouds show a steady increase in \xco at all \nhd column densities \citep{2011MNRAS.412.1686S}, at variance with the declining trends found in Fig. \ref{fig:XCOfactors}. 

We have also compared the mean visual extinction, \avavgco, mean CO intensity, $\overline{W_{\rm CO}}$, and \xco factor in each cloud with the simulation results of \citet{2011MNRAS.412..337G}. The $X_{\rm CO}(\overline{W_{\rm CO}})$ trend seen in Fig. \ref{fig:XCOfactors} roughly compares in slope with the simulation expectations, but not in absolute values, even though the efficiency of the radiation screening compares well in the observations and simulations. We show in Sect. \ref{sec:opacities} that the \anh ratio measured in the clouds compares well with the \cite{1978ApJ...224..132B} value adopted in the simulations. For a given \avavgco in a cloud, the simulations under-predict the mean \wcoavg intensity by a factor 20, and overpredict \xco by a factor 40. The comparison should be taken with care because both the observation and simulation samples are sparse, because the observations span a smaller dynamical range in \wcoavg than the simulations, and because the \wcoavg and \avavgco averages depend on the threshold applied to define CO cloud edges, which differs in both datasets. Nevertheless, the comparison highlights that, for initial volume densities $\leq 300$~cm$^{-3}$ applicable to the set of observed clouds, the simulations produce underluminous CO clouds at low densities compared to the observations. This was noted by \citet{2012AeA...544A..22L}. The CO deficit in their simulation reached a factor of ten in column density in the peripheral regions exposed to the UV radiation, where \nhd $\lesssim 2\times 10^{20}$ cm\msq. At larger column densities, the simulations and data compare much better. So, there seems to be a general problem with our understanding of chemistry at low gas density in the photo-dominated regions (PDR), either in the treatment of the UV attenuation or in the absence of warm chemistry driven by intermittent energy deposition in regions where the interstellar turbulence is dissipated \citep{2014AeA...570A..27G}.


The list of historical \xco measurements given in Table \ref{tab:Xco_history} and the values we are adding with the present sample of anti-centre clouds confirm a recurrent discrepancy that exists between the mean \xco factors measured with parsec resolution in nearby clouds and the \xco averages obtained at the scale of spiral arms (Local and Perseus arms) or of the Galactic disk. The former are two to three times lower than the large-scale values, which are close to 2 $\times$\xcounit (see Table \ref{tab:Xco_history}). 
Simulations of molecular clouds have warned that \xco factors can vary by up to a factor of five in the Milky Way because of the complexity of the overlap and blending in position and velocity of the CO clouds emitting along the lines of sight \citep{2011MNRAS.415.3253S}. \cite{2014MNRAS.441.1628S} have estimated an average value of 2.2 $\times$ \xcounit, close to the large-scale \g-ray measurements, when including all regions where CO intensities are potentially observable (i.e. wherever $>$0.1 K km s$^{-1}$).
But the spatial distribution of \hd column densities and CO intensities in their Galactic simulation implies significant deviations from this average, depending for instance on the location of cloud complexes with respect to the spiral arms.

Because of intrinsic \xco gradients across clouds,  \xco averages in large-scale observations respond to the mix of clouds in different states present in the region under study.  At least three potential reasons can explain why the large-scale \xco factor can exceed the local values. A first possibility is that a large fraction of the molecular mass lies in diffuse, gravitationally unbound clouds where \xco factors can reach very large values of 10$^{21-22}$ cm$^{-2}$ K$^{-1}$ km$^{-1}$ s \citep{2006MNRAS.371.1865B,2011MNRAS.412..337G}. A second possibility is that the average is driven by a large number of giant molecular clouds which are expected to exhibit \xco values of a few \xcounit because the gas is predominantly in the optically-thick CO regime.
A third influence is the level of shearing in the clouds. For instance, clouds sheared out by Galactic differential rotation are more exposed to photo-dissociation and should exhibit larger \xco values \citep{2014MNRAS.441.1628S}.

In parallel, the change in \xco estimates with scale may be due to determination biases induced by the sampling resolution in the gas maps, such as the increased difficulty at large distance to separate the clumpy CNM clouds and the DNM envelopes from their CO-bright phase. Misattributing a small part of the dense atomic gas or diffuse \hd would significantly bias the \xco factor upward. For instance, confusing 10--20\% of the CNM with the CO-related signal could explain the observed change in \xco by a factor of two between the local ISM and spiral arms \citep{2015AeA...582A..31A}. In nearby galaxies, \citet{2013ApJ...777....5S} have indeed measured a systematically larger \xco factor in highly inclined galaxies than in face-on ones where the pile-up along sight lines is reduced. 
More tests are needed to disentangle the origin of the \xco changes with scale. With larger photon statistics at high \g-ray energies, we should be able to investigate how the angular resolution and confusion between gas phases affect the calibration of \xco beyond the solar neighbourhood.

\begin{figure}[!h]
  \includegraphics[scale=0.37]{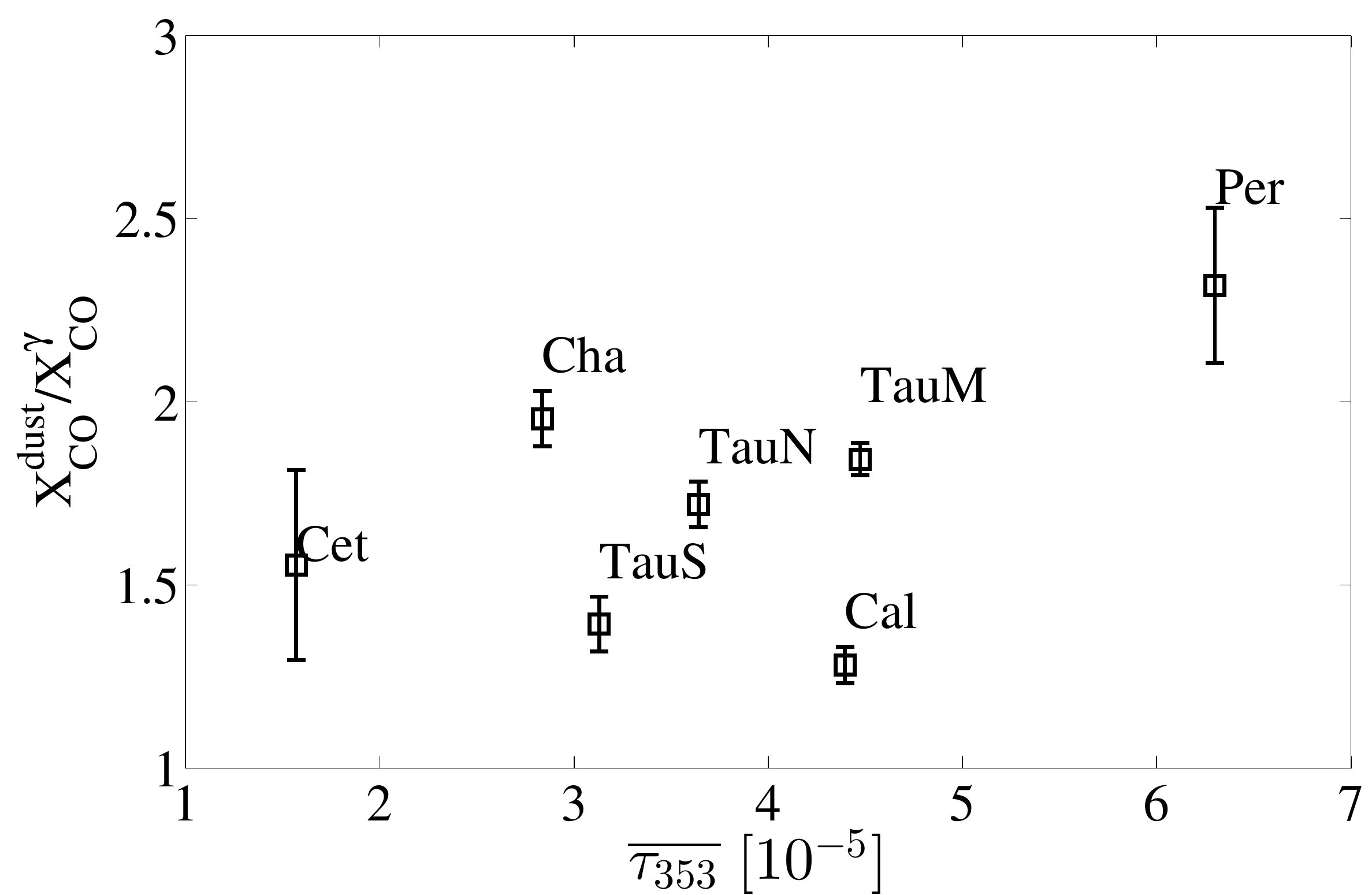}
  \caption{Ratio of \xco measurements with dust and \g rays in clouds of different average dust optical depth $\overline{\tau_{353}}$.}
   \label{fig:XCOfactors_dustgam}
\end{figure}

\subsection{X$_{\rm{CO}}$ measurements with dust} 

The results of the dust model provide an independent measure of \xco. Figure \ref{fig:XCOfactors_dustgam} shows that the dust-derived \xco factors are systematically 30\% to 130\% larger than the \g-ray values. Systematic discrepancies had already been noted in previous studies \citep{2011AeA...536A..19P,2013ARAeA..51..207B}, but on the basis of different \hi and CO surveys, different correlation methods, and different angular resolutions. In the present study, as in \cite{2015AeA...582A..31A}, the dust and \g-ray models are fitted to the same \nhi and \wco maps, at the same resolution. \citet{2013ARAeA..51..207B} proposed that the inclusion of the DNM gas in \g-ray analyses and its absence in dust analyses could explain the lower \xco values obtained in \g rays, but, by construction, the inclusion of the DNM map cannot bias the \xco estimate \citep{2005Sci...307.1292G,2015AeA...582A..31A}. Furthermore, the recent analyses include the DNM in both the dust and \g-ray fits and we have verified in this work that the \xco values change by less than 20\% when the DNM phase is added or not in the fits. 

Alternatively, the standard derivation of the \xco factor as $X_{\rm{CO}\tau}= y_{\rm{CO}}/(2y_{\rm{H_{I}}})$ is based on a uniform dust opacity, \opa, across the gas phases. A change in dust emission properties, in particular an increase in \opa at large gas column densities, can significantly bias the \xco factor upward \citep{2015AeA...582A..31A}. We show in Sect. \ref{sec:dust_evol} evidence for such a departure from linearity in dense CO regions where \opa can increase by a factor of up to six due to grain evolution compared to their properties in the atomic gas. Outside the densest molecular parts (brightest in CO), the molecular hydrogen is more diffuse and the increase in \opa is more modest, by a factor ranging between 1.2 and about 2, so the resulting bias on the derivation of the average value of \xco across a cloud is less than a factor of two.

Dust reddening and extinction maps have also been used to evaluate \xco in nearby clouds. We find values in close agreement with the extinction study of \citet{2015MNRAS.448.2187C} in the main Taurus and Perseus clouds (their Taurus and Perseus E1 values). The present Taurus South complex overlaps their Taurus E1, E2, and E3 structures with respective \xco factors of $(0.84 \pm 0.01)$, $(1.41 \pm 0.02)$, and  $(1.69 \pm 0.04) \times$~\xcounit bracketing our result in Taurus South. 

Using the 2MASS extinction map of \citet{2013PASJ...65...31D}, \citet{2012AeA...543A.103P} found \xco $=(2.27\pm 0.9) \times$~\xcounit in the Taurus-Perseus region, without decomposing the different clouds. Their \anh estimate of $5.33\times$ \anhunit in this region is 20\% lower than the all-sky $|b|>10^{\circ}$ average, but this difference is not sufficient to reconcile their measurement based on dust reddening with our values based on dust emission. The determination of \anh in the \hi phase directly impacts the derivation of \xco in the embedded molecular phase. The different methods used to infer \av from stellar data yield systematic differences as large as 1 magnitude at low \av, toward the atomic and diffuse ISM \citep{2013PASJ...65...31D,2015ApJ...810...25G,2016AeA...585A..38J}, so further analyses are necessary to evaluate their impact on \xco measurements.

\citet{2010ApJ...721..686P} have used 2MASS data and the NICER method to map the dust reddening and to derive \xco near $2.1 \times$~\xcounit in the Taurus Main complex. In this case, the difference with our result can be explained because they did not separate the extinction related to \hi along the CO sight lines, but they attributed all the dust to the molecular phase. In our analyses, we separate the dust or \g rays associated with each phase along all sight lines.


\section{Dust evolution}\label{sec:dust}

\subsection{Average dust opacities per gas phase}
\label{sec:opacities}

\input{AvNH_IG.txt}

Our analyses yield average dust properties per H atom in each gas phase. The best-fit y$_{\rm{HI}}$ coefficients directly give the average dust opacities, \opavghi, in the \hi structures. We have used the \g-ray estimates of the \xco factors to derive the average dust opacities in the CO-bright phase of each complex, \opavgco~$= y_{\rm{CO}}/(2 X_{\rm{CO} \gamma})$. The results are listed in Table \ref{tab:dustNH} together with similar measurements in the Chamaeleon clouds \citep{2015AeA...582A..31A}. To ease the comparison with other dust reddening measurements, we have converted the dust opacities to \nhebv ratios using the $E({\rm B-V})/\tau_{353}$ slope of $(1.49 \pm 0.03)\times 10^4$ found in a large correlation study between \taunu and reddening measurements toward 53 399 quasars \citep{2014AeA...571A..11P}. 
We did not attempt to estimate dust opacities in the ionised gas because of the large temperature gradients which impact the \taunu derivation toward the \hii regions, with uncertainties in \taunu doubling near the hotspots.

\begin{figure}[!t]
  \centering      
  \hspace{0.5cm}          
  \includegraphics[scale=0.54]{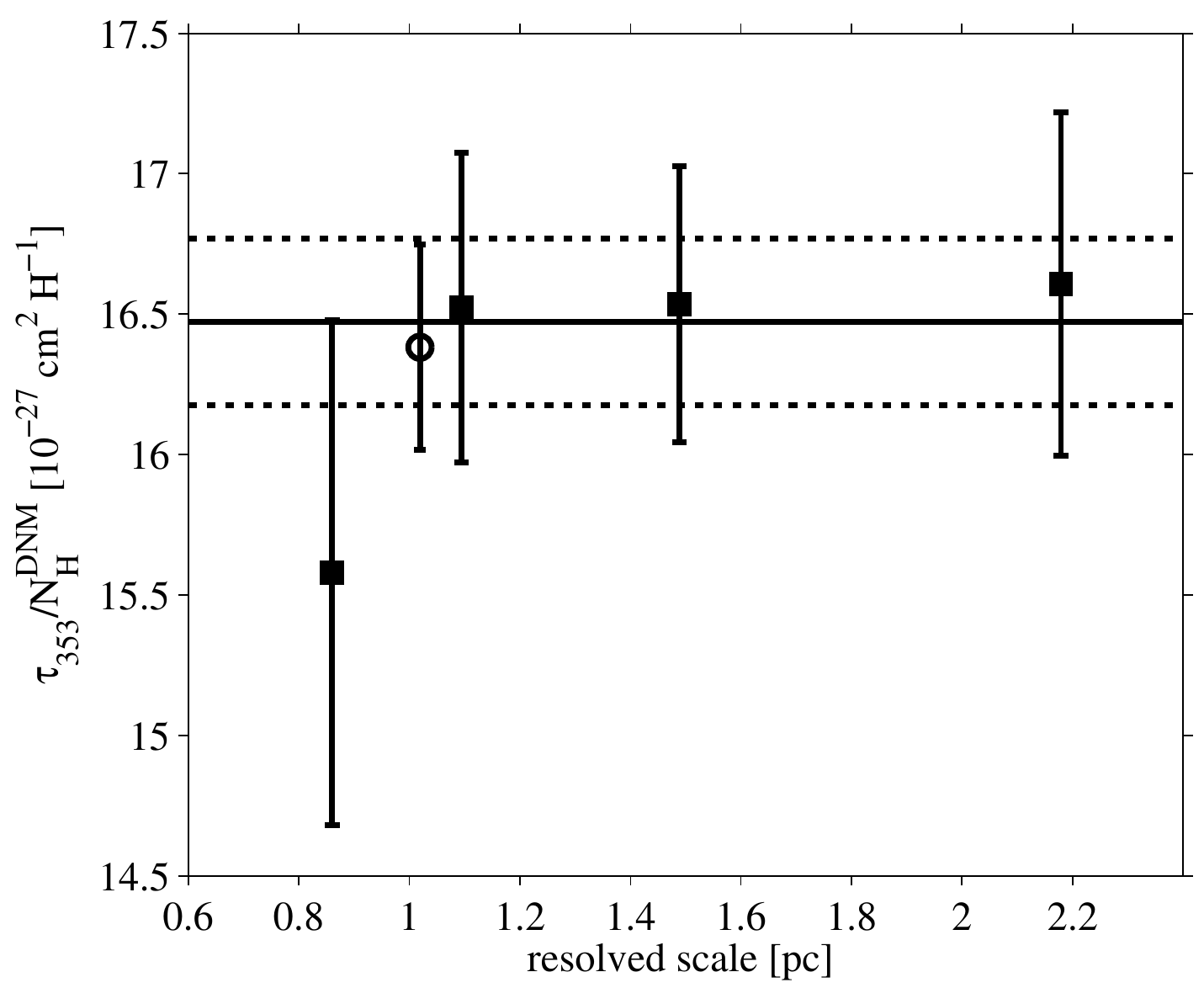}
  \hspace{0.5cm}
  \caption{Average dust opacities in the DNM measured in \g rays for different linear
resolutions in the clouds. The open circle marks the measurement in the 0.4--100 GeV energy band,  in close agreement with the weighted average of the four independent energy bands (black line) and its $\pm 1 \sigma$ errors (dashed lines).}
   \label{fig:res_scale_dust}
\end{figure} 

\begin{figure}[!h]
  \centering     
  \hspace{0.5cm}            
  \includegraphics[scale=0.5]{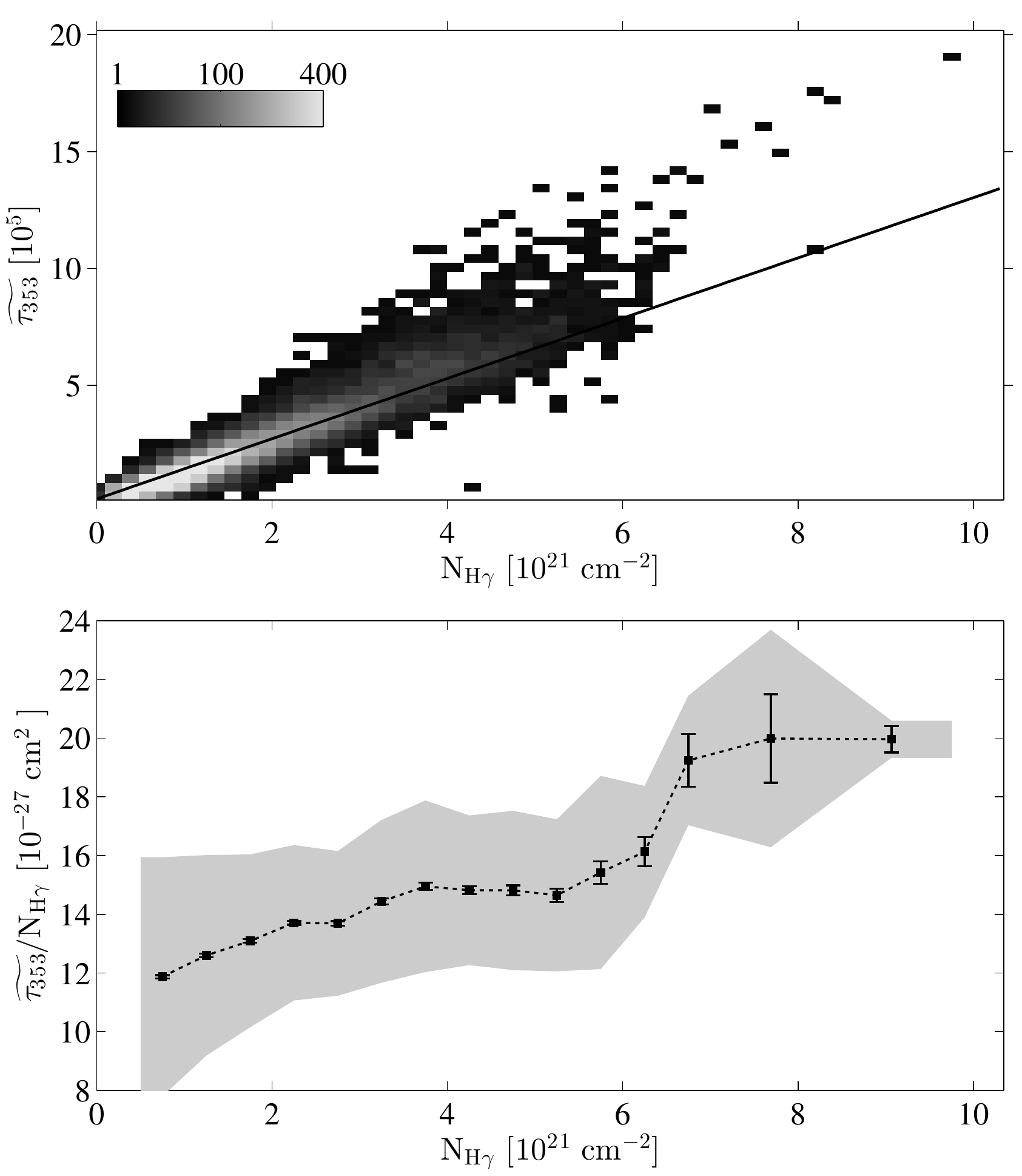}
  \hspace{0.5cm}
  \caption{ Upper : 2D histogram of the correlation between the total gas column density, \nhgam, measured with the 0.4--100 GeV interstellar \g rays, and the dust optical depth at 353 GHz, convolved with the LAT response for an interstellar spectrum. The maps were sampled at a 0\fdg375 resolution. Lower : evolution of the dust opacity in \nh bins. The error bars give the standard errors of the means and the grey band gives the standard deviation of the opacities in each bin.}
   \label{fig:dustVSgam}
 \end{figure}

The correlation existing between the \g-ray intensity and dust optical depth in the DNM yields an important measure of the average dust opacity in this phase. The value \opavgdnm~$=q_{\rm{HI}}/q_{\rm{DNM}}$ assumes a uniform CR flux across the \hi and DNM phases, an assumption that is corroborated by the fact that both phases exhibit comparable \g-ray emissivity spectra and moderate gas volume densities. According to Fig. \ref{fig:res_scale_dust}, we find no opacity changes at parsec scales in the extended DNM structures. The average, \opavgdnm~$= (16.5 \pm 0.3)\times 10^{-27}$~cm$^2$, compares very well with the value of $(17.2 \pm 0.5)\times 10^{-27}$~cm$^2$ obtained in the DNM surrounding the Chamaeleon clouds \citep{2015AeA...582A..31A}.

Using optically thin \hi, \cite{2014AeA...571A..11P} found marked gradients in dust opacity across the sky, with values ranging from 6.6 to about $11\times$\opaunit in the atomic gas, and a high-latitude value of $(7.0 \pm 2.0)\times$\opaunit in good agreement with the opacity of $(7.1\pm0.6)\times$\opaunit derived in high-latitude cirruses with a different analysis method \citep{2014AeA...566A..55P}. The opacities continued to increase in the CO-bright phase, up to $18\times$\opaunit for an \xco factor of \xcounit. Our results confirm that dust opacities vary from cloud to cloud and they can reach values well in excess of the high-latitude value, both in the atomic and molecular phase. 

\begin{figure*}
  \centering 
  \includegraphics[scale=0.7]{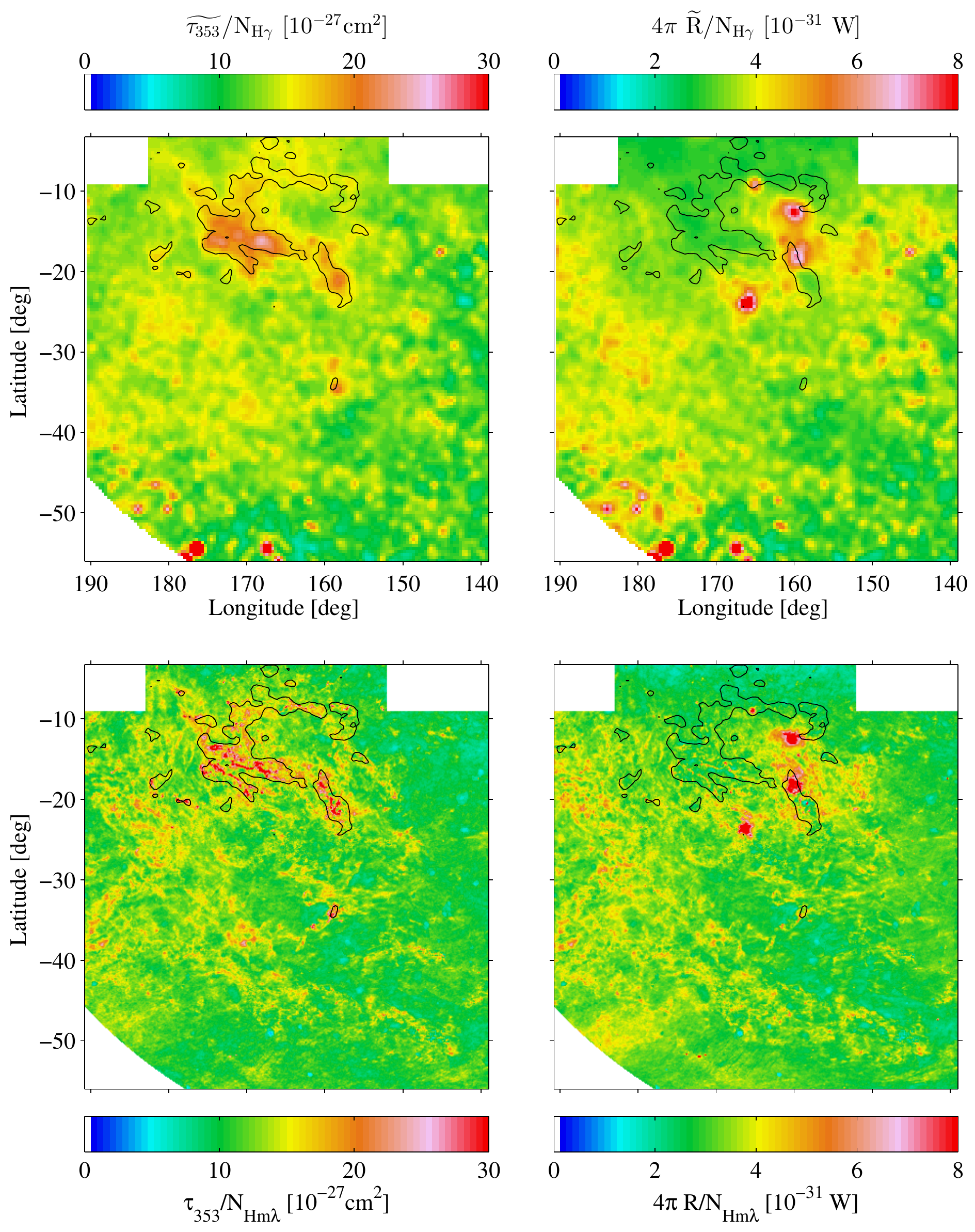}
  \caption{ Spatial variations of the dust opacities (left) and specific power (right) with the total gas measured by \nhgam at 0\fdg375 resolution (top) and by \nhlam at 0\fdg125 resolution (bottom). The tilded quantities are convolved with the LAT response for an interstellar spectrum. The black contours outline the shape of the CO clouds at the 7 K km s$^{-1}$ level chosen to separate DNM and \cosat components.}
   \label{fig:dustgamComp}
\end{figure*}  

  \begin{figure*}
  \centering 
  \hspace{0.4cm}               
  \includegraphics[scale=0.37]{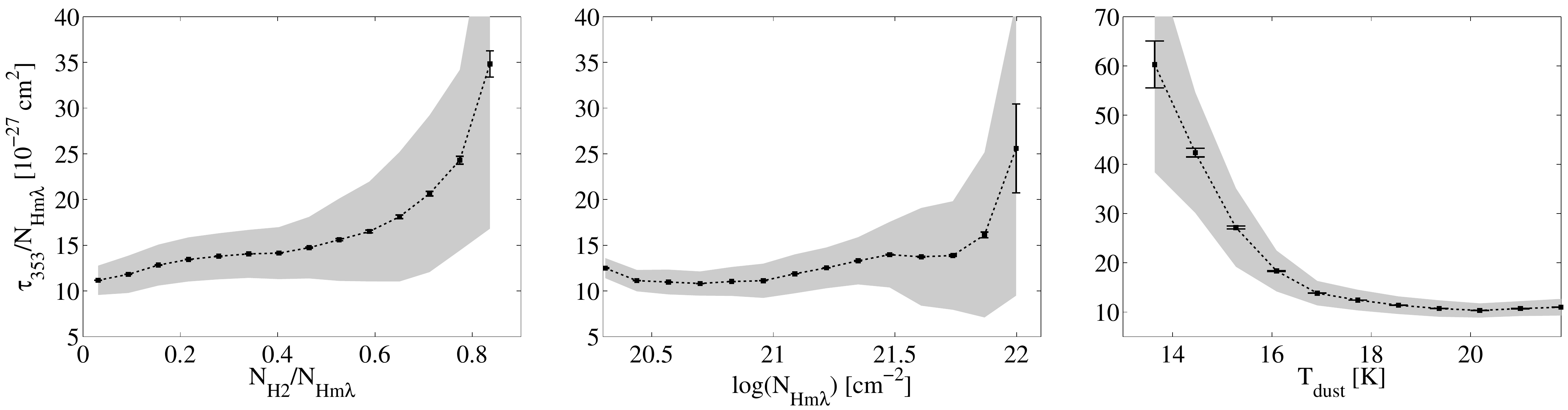}
  \caption{\small Evolution of the dust opacities as a function of the molecular fraction in the gas column (left), the total gas column density \nh (middle) and the dust color temperature (right). To estimate the molecular fraction, the DNM is assumed to be 50\% molecular. The error bars give the standard errors of the means and the grey band gives the standard deviation of the opacities in each bin.}
   \label{fig:tauVSfh2}
 \end{figure*}
 
We find 50\% to 135\% larger opacities in the anticentre \hi clouds than in the more diffuse \hi cirruses, even though we have corrected the \nhi column densities for a spin temperature of 400~K. The systematic uncertainties on \opa due to the \hi optical depth correction are listed in Table \ref{tab:dustNH} as the second set of errors. Given the likelihood variations shown in Fig. \ref{fig:dlnL_Tpsin}, we have taken the standard deviation in the set of measurements as a function of spin temperature for each parameter to represent such an uncertainty. Considerations of those uncertainties cannot reconcile the individual cloud measurements and the cirrus value, even for the rather tenuous \hi clouds in Cetus and Taurus North. The cloud-to-cloud dispersion indicates environmental effects beyond the latitude dependence noted by \cite{2014ApJ...783...17L} in the atomic gas. The \opavghi opacities listed in Table \ref{tab:dustNH} do not relate to the \hi mass of the corresponding clouds. The 15\% dispersion in this sample rather relates to the mean visual extinction of the \hi cloud and to the surface fraction of large and small \nhi column densities across the cloud. We present in the next section evidence for gradual opacity changes as the gas density increases that explain why the average opacity in an \hi cloud depends on its structure. The observation of opacity variations within the atomic phase at column densities below $10^{21}$ cm\msq challenges the current dust evolution models discussed in the next section.

We find systematically larger mean opacities in the CO-bright phase than in the \hi phase, thereby confirming the continued increase in \opa at large \hd column densities. The latter has been evaluated independently in each cloud (independent \xco factors), so the cloud-to-cloud changes in \opavgco that \cite{2014AeA...571A..11P} have noted across the sky are not due to their use of a uniform \xco factor. In Table \ref{tab:dustNH}, the mean opacity increases by 30\% to 100\% between the \hi and CO-bright phases, independent of the \hd mass locked in the cloud. The dispersion in \opavgco rather relates to the surface fraction of dense regions with large \wco intensities, $\rm{SF}_{dense}$, in the averaging. To study environmental changes in \opa within molecular clouds, we must take into account the added complication of spatial gradients in \xco. 


\subsection{Environmental changes in dust opacity}
\label{sec:dust_evol}

In order to follow changes in dust opacity across the clouds, we have built two maps of the total \nh column density in the region. The first one, \nhgam, takes advantage of the CR interactions with gas in all chemical forms and thermodynamical states. The \g-ray intensity from the gas has been obtained from the LAT data in the overall energy band after subtraction of the \g-ray counts unrelated to gas in the best-fit model. In order to reduce the Poisson noise and get more robust photon statistics, we have degraded the resolution to 0\fdg375.
We have converted this \g-ray intensity into \nh using the average emission rate, $\overline{q}_{\rm HI}$, obtained in the atomic phase in the region. The second map, \nhlam, uses the higher-resolution information from the \hi, CO, and dust data, and the mass scaling provided by the best-fit \g-ray model (uniform \hi spin temperature of 400~K, \xcoG factors in the clouds, \g-ray emissivity in the DNM, ionised, and \cosat gas as in the \hi). 
The opacity distribution at the smallest angular scales should be considered with care because the scaling factors used to construct \nhlam have been derived at the scale of a whole cloud.

Figure \ref{fig:dustVSgam} shows the 2D histogram of the correlation between the sightline integrals in dust optical depth and in \nhgam column density. The distribution deviates from linearity above about $4\times 10^{21}$~cm\msq. The opacity maps, shown in Fig. \ref{fig:dustgamComp}, 
indicate that \opa evolves progressively within the clouds, especially when the medium becomes molecular, in the diffuse DNM and even more so in the CO-bright parts. The increase in opacity does not relate to an increase in the specific power, \spw, radiated by the grains, so the rise in opacity cannot be attributed to a larger heating rate (if we consider that the large grains are in thermal equilibrium to equate the radiated and absorbed powers). The largest opacities rather correspond to a ${\sim}30$\% lower heating rate in the well-shielded CO regions.

Thanks to the total gas tracing capability of the CRs and to the complementarity of the \nhgam and \nhlam maps, it is very unlikely that opacity variations as large as those seen in Fig. \ref{fig:dustgamComp} are caused by large deficits in the gas column densities. Our results  give strong support to earlier indications that the big grains in the CO filaments of the Taurus cloud have larger sub-mm emissivities than in the more diffuse media. The previous studies noted opacity changes 
by a factor ranging from ${\sim}2$ \citep{2009ApJ...701.1450F,2011AeA...536A..25P,2013AeA...559A.133Y} to $3.4^{+0.3}_{-0.7}$ \citep{2003AeA...398..551S} above the diffuse-ISM value. In the present analysis, we find opacity enhancements exceeding a factor of three and reaching a factor of six toward the CO clouds, even though we provide a measure of the additional \hd gas that is not linearly traced by \wco in the \cosat directions. The largest opacities do not rise as high in the Taurus clouds as they do in the Chamaeleon ones \citep{2015AeA...582A..31A}, but a quantitative comparison would require the separation of the DNM and \cosat components in a new analysis of the Chamaeleon region. We also find that the opacity starts to increase in the DNM, therefore at lower gas densities than the few thousand per cm$^3$ sampled in CO. The grain opacity at the \hi--\hd transition compares very well with that in the DNM surrounding the Chamaeleon clouds \citep{2015AeA...582A..31A}. 
 \begin{figure}[!t]
 
  \centering 
  \hspace{0.4cm}               
  \includegraphics[scale=0.56]{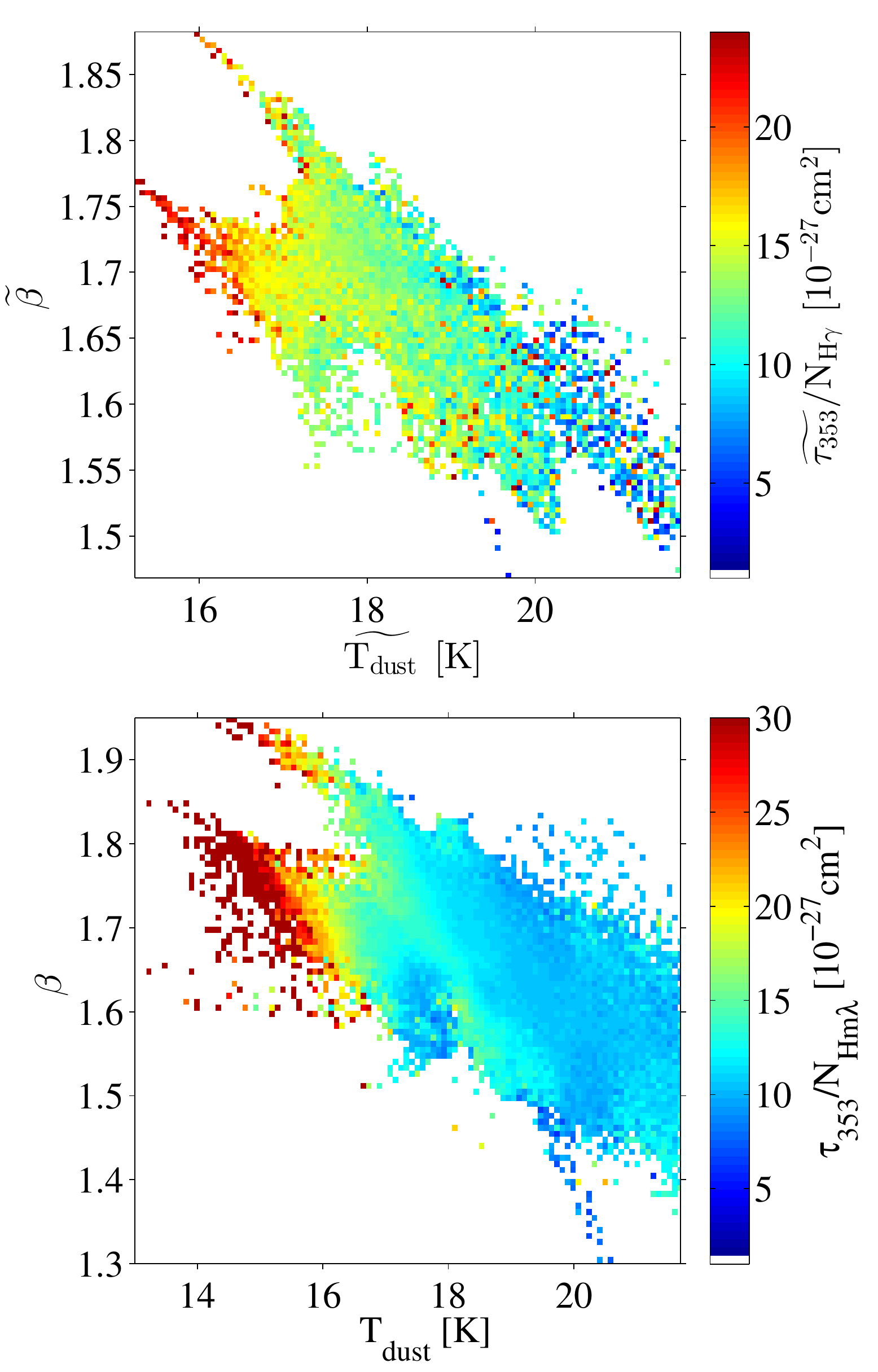}
  \hspace{0.4cm}
  \caption{\small Evolution of the dust opacities with the total gas measured by \nhgam at 0\fdg375 resolution (top) and by \nhlam at 0\fdg125 resolution (bottom) as a function of the dust color temperature (x-axis) and spectral $\beta$ index (y-axis)}
   \label{fig:taunh_beta_T}
 \end{figure}
 
In order to determine how the opacity gradient depends steeply on the ambient ISM, Fig. \ref{fig:tauVSfh2} shows how the opacity varies with the molecular fraction in the gas column, f$_{\rm{H}_2}$, with the total \nhlam column density as traced by the multi-wavelength data, and with the dust colour temperature. We observe a gradual increase as the gas becomes molecular and a steeper change at \nh above $5\times 10^{21}$~cm\msq due to two factors: change in molecular fraction and in grain properties in the molecular gas. The amplitude of the gradient is best traced by the drop in grain temperature. Figure \ref{fig:dustgamComp} further shows that the ambient molecular gas density is not the only factor of evolution because we detect notable cloud-to-cloud differences in the opacity gradients. Compare for instance the large opacities that often exceed $25\times$\opaunit along the Main Taurus and Perseus CO clouds with the 50$\%$ lower values obtained toward comparable \wco intensities near both ends (in longitude) of the California cloud. 

Several studies have noted that the opacity enhancement occurred together with a reduction of the IR emission from polycyclic aromatic hydrocarbons (PAHs) and from very small grains stochastically heated by the ambient ISRF \citep[e.g.,][]{2003AeA...398..551S,2009ApJ...701.1450F,2013AeA...559A.133Y}. Grain evolution in dense media was invoked to explain the radiation changes \citep[see the previous references and][]{2012ApJ...751...28M,2015AeA...579A..15K}. It involves accretion of aliphatic-rich carbonaceous mantles from the gas phase, grain coagulation into fluffy aggregates, and the formation of an ice mantle on the surface of the aggregates in the coldest regions. Recent theoretical studies such as \cite{2015AeA...579A..15K} suggest that the spectral changes in $\beta$ relate to gas density-dependent processes and that the variations in opacity, spectral index $\beta$, and colour temperature are closely connected. Figure \ref{fig:taunh_beta_T} shows the quantitative relation between those three parameters in the observations. The data corroborate the prediction that the opacity should increase for $\beta$ above 1.5 and temperature below 17~K. According to this model, the directions with temperatures near 16--18~K, $\beta \ge 1.5$, and $\tau_{353}/N_{\rm Hm\lambda}$ values about twice as large as the diffuse-ISM value of ${\sim}7 \times$\opaunit would be rich in grains with carbonaceous mantles. The mantle accretion would therefore occur primarily in the DNM. Aggregates would dominate the directions below 16~K, at $1.6 \lesssim \beta \lesssim 1.8$, and where \opa is at least four times above the diffuse-ISM value. Such conditions are found inside the sampled CO clouds. Ice coating of the aggregates would explain the largest opacity values in Fig. \ref{fig:taunh_beta_T}, at the lowest temperatures and largest spectral indices, $\beta$.

\section{Conclusions}


We have analysed the gas, dust and CR content of several nearby anti-centre clouds including the Cetus, Taurus, Auriga, Perseus, and California clouds. We have performed an iterative fit of the total gas column density as traced by $\gamma$ rays and the dust optical depth at 353 GHz.
We have modelled both tracers as linear combinations of the gas column densities traced by \hi, $^{12}$CO, and free-free emissions, and we have extracted the morphology and 
column density of the DNM from the joint \g-ray and dust data. 
We have verified the robustness of our set of parameters through the use of jackknife tests; these parameters have enabled us to constrain the properties of the ISM in 
these clouds.

The main results are summarised as follows. 

\begin{itemize}
\item \textbf{On the cosmic rays:}
The \g-ray emissivity of the gas in the analysed clouds has the same energy spectrum as in other clouds of the local ISM and we find no dependence of their average \g-ray emissivity with Galactocentric radius (i.e., distance from the local spiral arm), nor with height above the Galactic plane. In the 0.4--100 GeV energy band and at the precision level of the current LAT observations, we find no evidence of CR exclusion or CR concentration in the clouds, up to the $^{12}$CO-bright molecular regions. 
\item \textbf{On the dark gas:}
The \g-ray and dust data jointly reveal significant amounts of gas in addition to that seen in \hi, free-free, and $^{12}$CO emissions. The diffuse large-scale structures are associated with the DNM at the transition between the atomic and molecular phases. They gather dense atomic hydrogen and diffuse \hd in unknown proportions, but with column densities equivalent to those found in the \hi and CO emitting parts. In the molecular phase, the \g rays and dust reveal filaments of dense gas in addition to that proportionality traced by \wco where the $^{12}$CO line intensities saturate. 

\item \textbf{On the \hii regions:}
The \hii regions, NGC 1499 and G156.6-18.5, are jointly detected in dust and \g-ray emission. The corresponding \g-ray flux cannot be attributed to the IC up-scattering of the bright stellar radiation by the local CR electrons. It is more likely caused by the hadronic interactions of the local CR nuclei in the ionised gas. Hence for NGC 1499, we have used the average \g-ray emissivity of the atomic gas in the region to infer a mean electron density of 4.3$\pm$0.6 cm$^{-3}$ for an electron temperature of 8000 K, in good agreement with the modelling of the free-free and H$\alpha$ emissions of the \hii region.

\item \textbf{On the \xco factors:}
We provide independent measurements of the \xco factor from the dust and \g-ray analyses in six different clouds, against the same \hi and CO data. As in the Chamaeleon complex, we find that the dust-derived values are systematically larger, by 30\% to 130\%, than the \g-ray estimates. The difference is likely due to chemical and structural evolution of the dust grains with increasing \nh. The \xco factors measured in \g rays range from $(1.04\pm0.05)\times$ \xcounit in Taurus South to $(0.44\pm0.04)\times$ \xcounit in Perseus. Together with the estimate found in the Chamaeleon clouds \citep{2015AeA...582A..31A}, these measurements indicate that \xco values tend to decrease with the average \wco intensity of a cloud, with the average CO line width, and with the surface fraction subtended by the brightest CO clumps. The more diffuse CO clouds therefore tend to have larger average \xco factors. Models of the formation and photodissociation of \hd and CO molecules predict a marked decline in \xco from the diffuse envelopes of molecular clouds to their dense cores. Hence the \xco average should qualitatively vary with the surface fraction of dense regions in the cloud structure and with the CO line width, as in our data. We find a modest change by typically a factor of two in \xco with cloud state, but our sample lacks a giant molecular cloud to extend the state range. The amplitude of these variations already limit the precision of molecular mass estimates based on the use of CO intensities and of a mean \xco conversion factor. We further note that the low \xco values measured here are at variance with the theoretical predictions from PDR chemistry for the moderate gas densities filling most of the volume of the observed clouds. 
The present measurements confirm a recurrent discrepancy by a factor of two to three between the \xco averages obtained in nearby clouds and at large scale in the Galaxy. Simulations predict \xco variations of this amplitude in the Galactic disc  owing to the variety of CO-line blending situations occurring along the lines of sight \citep{2011MNRAS.415.3253S,2014MNRAS.441.1628S}. A dominance of diffuse, unbound or sheared clouds, or of giant molecular clouds, can drive the \xco factor of the ensemble to large values.  Additionally, the increased level of cross-correlation at large distance between the CNM, DNM, and CO-bright phases of the ISM could bias the large-scale \xco determination upward.

\item \textbf{On the dust opacities:}
The dust opacity at 353 GHz, \opa, appears to rise by a factor of three from low column densities in the atomic gas (about $10^{-26}$~cm$^2$) to cold grains in molecular gas at \nh$\gtrsim 5\times 10^{21}$~cm\msq. The rise can reach a factor of six at very low dust temperatures below 14 K. The amplitude of the rise is comparable to the variations observed in the Chamaeleon clouds \citep{2015AeA...582A..31A}. As the observed specific power radiated by the grains decreases in the cold molecular regions the changes cannot be attributed to a larger heating rate, but they are more likely caused by a chemical or structural change in the grains. The magnitude of the rise severely limits the use of the thermal emission of the large dust grains to trace the total gas. The linear regime is limited to \nh $< 3\times 10^{21}$~cm\msq in these clouds. The confirmation of large opacity variations across clouds directly impacts the gas mass estimates inferred from dust emission at sub-mm and mm wavelengths to derive star-forming efficiencies in the Galaxy and in external galaxies. We quantify the coupled changes in grain temperature, opacity, and $\beta$ index of the thermal radiation to help model the grain evolution.
\end{itemize}

In a second paper we will study how the DNM and CO$_{\rm sat}$ gas relate to the \hi-bright, $^{12}$CO-bright, and $^{13}$CO-bright phases. 
We will derive the masses of the DNM and of the CO$_{\rm sat}$ phase for a sample of substructures in the Cetus, Taurus, California and Chameleon clouds, while taking care to avoid regions of cloud confusion along sightlines.
We will follow the evolution of the CO-dark \hd fraction both within a cloud and from cloud to cloud  and we will place our results in the context of other observational and theoretical studies. 

In a subsequent study, we will confront the gas, \g-ray, and dust distributions in the nearby clouds using dust column densities inferred from stellar reddening. 
The joint comparison of dust emission and extinction data to other gas tracers will provide additional information on the relation between the total gas and dust, including the abundant dark neutral medium, and will help to better constrain the evolution of dust properties, in particular with regard to potential variations of the extinction curve \citep{2016ApJ...821...78S}.

Adding more data points to the trends shown in Fig. \ref{fig:XCOfactors} is essential to confirm that the average \xco factor in a cloud tends to decrease for increasingly dense CO molecular clouds, in response to different \xco gradients spanning each cloud \citep{2006MNRAS.371.1865B,2011MNRAS.412..337G}. Particular attention should be paid to the use of the same method to derive a consistent set of \xco values, and to the addition of clouds with little confusion between the different gas phases to ensure a precise separation of the total gas tracers in each phase. The current \g-ray analyses cannot yield \xco maps across individual clouds. The variations in dust opacity with increasing gas column density prevent the use of dust emission maps to probe \xco gradients in nearby clouds. In this context, it is essential to increase the number of clouds with measured average \xco values and to confront the observed \xco trends with PDR models in order to interpret the slope and dispersion of the trends presented here. Confirming that \xco factors vary from cloud to cloud within the local ISM because of their different dynamical and chemical structures stresses that large-scale \xco averages can vary because of a different mix of cloud states in different Galactic regions, in addition to the Galaxy-wide \xco gradients expected from metallicity and ISRF gradients.  For instance, it opens the way for observable differences in \xco between ensembles of clouds inside and outside of the spiral arms since the inter-arm clouds that have been sheared after their passage through an arm are more susceptible to photo-dissociation, thus more prone to having large \xco averages \citep{2014MNRAS.441.1628S}. 

\begin{acknowledgements}
We thank the referee for an interesting discussion. 
The \textit{Fermi} LAT Collaboration acknowledges generous ongoing support
from a number of agencies and institutes that have supported both the
development and the operation of the LAT as well as scientific data analysis.
These include the National Aeronautics and Space Administration and the
Department of Energy in the United States, the Commissariat \`a l'Energie Atomique
and the Centre National de la Recherche Scientifique / Institut National de Physique
Nucl\'eaire et de Physique des Particules in France, the Agenzia Spaziale Italiana
and the Istituto Nazionale di Fisica Nucleare in Italy, the Ministry of Education,
Culture, Sports, Science and Technology (MEXT), High Energy Accelerator Research
Organization (KEK) and Japan Aerospace Exploration Agency (JAXA) in Japan, and
the K.~A.~Wallenberg Foundation, the Swedish Research Council and the
Swedish National Space Board in Sweden. Additional support for science analysis during the operations phase is gratefully acknowledged from the Istituto Nazionale di Astrofisica in Italy and the Centre National d'\'Etudes Spatiales in France.
The authors acknowledge the support of the Agence Nationale de la Recherche (ANR) under award number STILISM ANR-12-BS05-0016. 
\end{acknowledgements}

\bibliographystyle{aa}
\bibliography{biblioG}

\appendix
\section{Component separation} 
\label{sec:AnnexCS}

We have decomposed the \hi and CO spectra into sets of individual lines as in \cite{2015AeA...582A..31A}. The method starts with the detection and localization in velocity of isolated lines or significant peaks in the spectra. Each HI and CO spectrum is then fitted by a sum of pseudo-Voigt line profiles centred on the centroids of the detected lines or peaks, within $\pm$2.7 to 3~km/s. The fits are generally good, with less than 15$\%$ difference between the observed and fitted integrals over all velocities. In order to preserve the observed photometry, we calculate the residuals between the observed and fitted spectra in each channel and we redistribute them among the fitted lines, proportionally to their intensity in this channel. We finally construct \nhi column-density and \wco intensity maps of a specific cloud by integrating the fitted lines that have their velocity centroid within a chosen velocity interval for each (\lt, \bt) direction. This method corrects for potential line spill-over from one velocity interval (i.e., cloud) to the next. To determine boundaries between clouds in the position-velocity space (\lt, \bt, v), we have used the velocity centroids of all the fitted lines and their density in the (\lt, \bt, v) space to search for line clusters (i.e., clouds). To do so, we have used the Density-Based Spatial Clustering of Applications with Noise (DBSCAN) algorithm \citep{1996DBSAN} to isolate arbitrarily-shaped clusters and to draw contours of minimum density around them.
The six \hi and CO regions used in this analysis have been delimited by following those curves of minimal number density  in (l, b, v) space. The cuts follow piecewise continuous lines in (b,v) over specific longitude intervals. The latter and the points used to define the broken lines are given in Table \ref{tab:CompLim}.

\input{VelSep.txt}

\section{Free-free emission in the 70 GHz data} 
\label{sec:AnnexFF}
In order to separate the free-free intensity in the \textit{Planck} LFI map at 70 GHz from non-gaseous emission components, we have proceeded as follows. We have first subtracted the CMB intensity as modelled at 70 GHz by \cite{2016AeA...591A..50B} from the joint analysis of \textit{WMAP} and \textit{Planck} observations. We have then taken advantage of the sparse spatial correlation between the thermal dust intensity $I_{70}$ at 70 GHz and the dust optical depth \taunu at 353 GHz in the wavelet domain to filter out the dust component. To do so, we have transformed both maps with eight scales of the B-spline wavelet with the \texttt{mr-transform} tool of \cite{1998AeAS..128..397S}. For each scale, we have obtained the best ($\chi^2$) linear regression between the wavelet distributions of $I_{70}$ and \taunu in the regions of both significant dust optical depth ($\tau_{353} > 5\,10^{-6}$) and low free-free intensity ($< 0.3$~mK$_{\rm CMB}$ at 22 GHz). We have used this regression to remove the part linearly correlated with \taunu from the 70 GHz transform at each scale. After this correction, we have applied the inverse transform to reconstruct the filtered map. We have then used the \texttt{mr-detect} tool with the same eight wavelet scales to detect and remove point sources. Only the first scale was used for source detection. The final map has been translated into units of specific intensity (Jy/sr) using the conversion coefficients\footnote{\url{http://wiki.cosmos.esa.int/planckpla2015/index.php/UC_CC_Tables#LFI_Unit_Conversion_Tables}} for the 70 GHz band. We have applied the colour correction interpolated for the -0.14 spectral index expected for free-free emission at 70 GHz.

\section{\hi spin temperature} 
\label{sec:AnnexT}

\begin{figure}[!h]
  \centering                
  \includegraphics[width=\hsize]{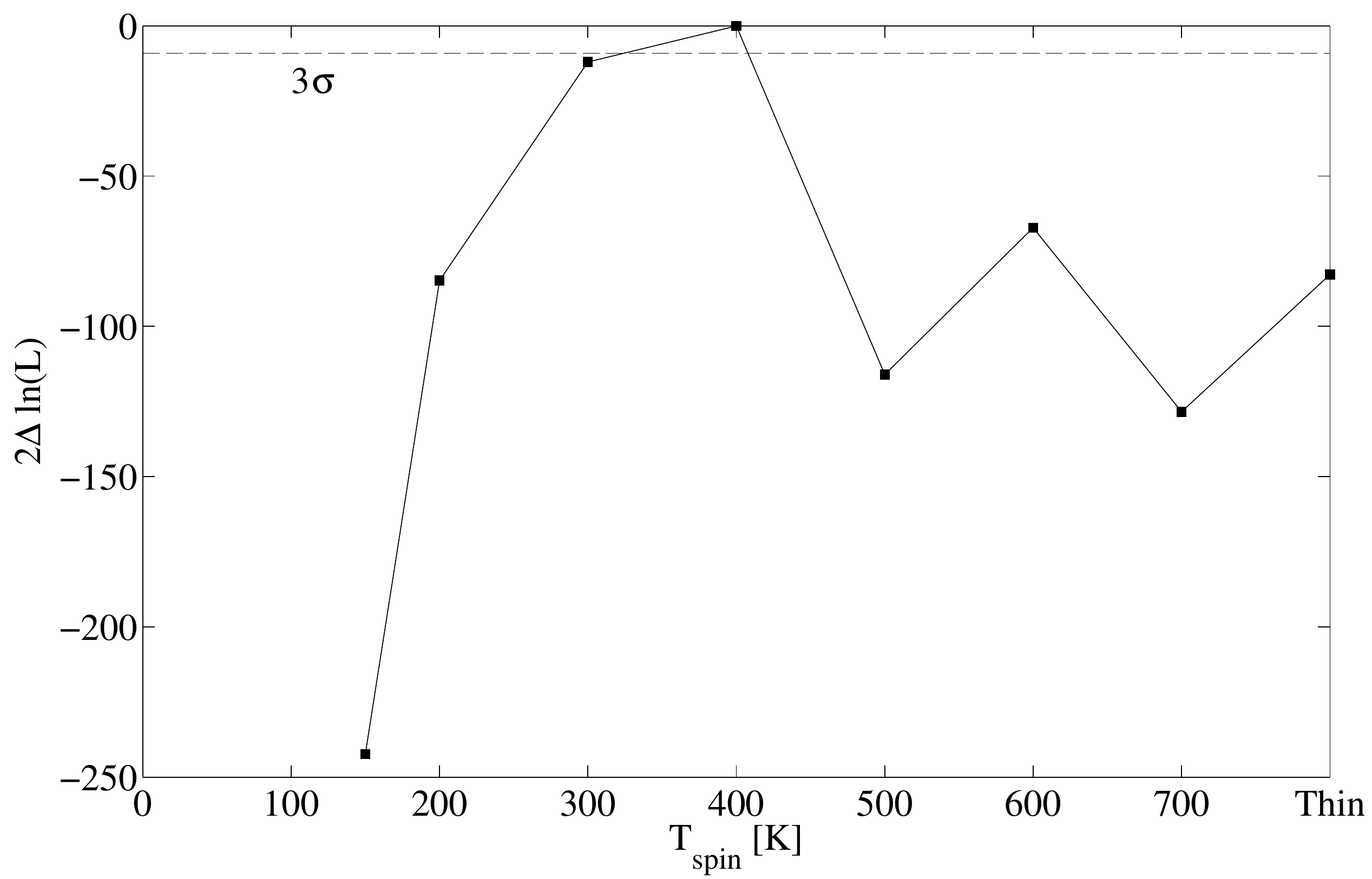}
  \caption{ Evolution of the log-likelihood ratio of the \g-ray fit in the 0.4-100 GeV band as a function of the \hi spin temperature used to calculate the \nhi column densities in the clouds.}
   \label{fig:dlnL_Tpsin}
\end{figure}

The evolution of the maximum log-likelihood of the \g-ray fit in the 10$^{2.6-5}$ MeV band can be used to constrain the choice of average \hi spin temperature that best matches the \nhi column density maps of the different clouds to the structure of the \g rays produced by CR interactions in the atomic gas. We find that the fit quality peaks near 400 K (Fig. \ref{fig:dlnL_Tpsin}). We have not attempted to use different spin temperatures for the different clouds present in the analysis region.

\section{Best fit coefficients of the $\gamma$-ray and dust models} 
\label{sec:AnnexFITs}
The best fit corresponds to an H$_{\rm I}$ spin temperature of 400 K.
For each parameter we first give the statistical uncertainty from the fit and jackknife tests, then the standard deviation obtained by varying the \hi spin temperature.
\input{ResFit.txt}

\section{Previous $\gamma$-ray measurements of the $X_{\rm CO}$ conversion factor}
\input{Xco_history.txt}

\end{document}

%% file: XCO_IG.txt
\renewcommand{\arraystretch}{1.7}
\setlength{\tabcolsep}{0.17cm}
\begin{table}
\caption{ $X_{\rm{CO}}$ factors in $10^{20}$ cm$^{-2}$K$^{-1}$km$^{-1}$s for the dust and $\gamma$-ray fits.}
\centering
\begin{tabular}{l|c|c}
 Cloud&$\tau_{353}$ &$\gamma$ rays \\ \hline 
Cetus & 1.58$\pm$0.08$\pm$0.07 & 1.01$\pm$0.16$\pm$0.07 \\ 
Taurus South & 1.44$\pm$0.04$\pm$0.04 & 1.04$\pm$0.05$\pm$0.05 \\ 
Taurus North & 1.55$\pm$0.03$\pm$0.05 & 0.90$\pm$0.03$\pm$0.05 \\ 
Taurus Main & 1.24$\pm$0.02$\pm$0.06 & 0.67$\pm$0.01$\pm$0.04 \\ 
California & 1.11$\pm$0.03$\pm$0.07 & 0.87$\pm$0.03$\pm$0.03 \\ 
Perseus & 1.03$\pm$0.05$\pm$0.12 & 0.44$\pm$0.04$\pm$0.04 \\ 
\end{tabular}
\tablefoot{The first uncertainties are statistical, the second give the standard deviations obtained by varying the \hi spin temperature.}

\label{tab:XCOfactors}
\end{table}

%% file: AvNH_IG.txt
\renewcommand{\arraystretch}{1.7}
\setlength{\tabcolsep}{0.19cm}
\begin{table*}
\caption{Average dust opacities, \opa, and $N_{\rm H}/E(B-V)$ ratios in the gas phases of the different clouds.}
\centering
\begin{tabular}{l|c|c| c|cc}
 Cloud &  \opavghi  & \opavgco & $\overline{\left[\frac{N_{\rm H}}{E({\rm B}-{\rm V})}\right]}_{\rm HI}$ &  $\overline{\left[\frac{N_{\rm H}}{E({\rm B}-{\rm V})}\right]}_{\rm CO}$ \\ 
& [$10^{-27}$ cm$^2$] & [$10^{-27}$ cm$^2$] & [$10^{21}$ cm$^{-2}$] & [$10^{21}$ cm$^{-2}$] \\  
\hline 
$\rm{Cet}$ & 10.3$\pm$0.1$\pm$0.2 & 17.9$\pm$4.1$\pm$1.2 & 6.54$\pm$0.19$\pm$0.15 & 3.75$\pm$0.94$\pm$0.25 \\ 
$\rm{TauS}$ & 13.7$\pm$0.1$\pm$0.4 & 18.3$\pm$1.5$\pm$1.1 & 4.91$\pm$0.12$\pm$0.15 & 3.67$\pm$0.38$\pm$0.22 \\ 
$\rm{TauN}$ & 11.0$\pm$0.1$\pm$0.3 & 18.8$\pm$1.0$\pm$1.0 & 6.12$\pm$0.16$\pm$0.18 & 3.56$\pm$0.27$\pm$0.19 \\ 
$\rm{TauM}$ & 14.2$\pm$0.1$\pm$0.6 & 27.0$\pm$1.0$\pm$1.5 & 4.73$\pm$0.13$\pm$0.20 & 2.48$\pm$0.14$\pm$0.14 \\ 
$\rm{Cal}$ & 13.7$\pm$0.2$\pm$0.5 & 17.7$\pm$1.0$\pm$0.7 & 4.91$\pm$0.16$\pm$0.17 & 3.80$\pm$0.28$\pm$0.15 \\ 
$\rm{Per}$ & 16.5$\pm$0.5$\pm$1.0 & 34.4$\pm$5.1$\pm$4.6 & 4.08$\pm$0.20$\pm$0.26 & 1.95$\pm$0.33$\pm$0.25 \\
$\rm{Cha}^a$ & $16.3 \pm 0.2^{+0}_{-0.8}$ & $32 \pm 1^{+0}_{-2}$ & $4.11 \pm 0.10^{+0.23}_{-0}$ & $2.1 \pm 0.2^{+0.1}_{-0}$ \\ 
$\rm{Cha-IVA}^a$ & $14.8 \pm 0.2^{+0}_{-0.6}$ & & $4.5 \pm 0.1^{+0.2}_{-0}$ 
\end{tabular}
\vspace{0.3cm}
\tablefoot{$^a$ from \cite{2015AeA...582A..31A}.\\
The first uncertainties are statistical, the second give the standard deviations obtained by varying the \hi spin temperature.}

\label{tab:dustNH}
\normalsize
\end{table*}

%% file: VelSep.txt
\begin{table*}[!h]
\caption{Limits of the \hi and CO clouds in longitude, latitude, and velocity }
\centering
\begin{tabular}{l|c |c |c |c |c |c }
\hline
Cloud & Cetus & Main Taurus & South Taurus & North Taurus & Perseus & California  \\ \hline
l$_{\rm{min}}$& 135\degree&135\degree &135\degree &135\degree & 155\degree & 151\fdg6\\ \hline
l$_{\rm{max}}$&195\degree & 195\degree& 195\degree& 195\degree& 161\fdg5 & 178\fdg8\\ \hline
(b , v) &-22\degree , -14 & 3\degree , 5.8&3\degree , 5.8&1\fdg4 , -2.3&-17\fdg8 , 8\fdg2&-3\fdg6 , 0.8\\
points &-30\degree , 16& -2\degree , 6.1 & -2\degree , 6.1 & -3\fdg6 , -3.6 & -17\fdg3 , 6 & -6\fdg5 , -1\\ 
for &-42\degree , -10&-5\fdg5 , 5.9&-3\degree , 6.1&-6\degree , -7.2&-18\fdg5 , 6.2&-7\degree , -0.25\\
lower	  &-61\degree , -100&-8\degree , 4.3&-3\fdg5 , 7.2&-10\fdg6 , -12.2&-22\fdg1 , 1.3&-8\fdg6 , 1\\
border	  & &-9\degree , 1.7&-6\degree , 14&-14\fdg1 , -11.8&-23\fdg9 , 4.4&-10\fdg1 , 0.6\\
	  & &-10\degree , 1.4&-19\degree , 14&-18\degree , -6& &-14\fdg1 , -2\\
	  & &-11\fdg5 , 3.&-21\fdg2 , 9.3&-20\degree , 14& &-14\fdg1 , -11.8\\
	  & &-13\degree , 3.1&-23\fdg9 , 4.4&-31\degree , -14& & \\
	  & &-14\fdg6 , 4.2&-25\degree , 6&-35\fdg5 , 6& &\\
	  & &-17\degree , 4.2&-34\fdg5 , 6&-42\degree , -7& &\\
	  & &-18\fdg5 , 6.2&-38\fdg6 , 2.8&-61\degree , -6& & \\
	  & &-22\degree , 1.3&-61\degree , 7.2& & &\\
	  & &-23\fdg9 , 4.4& & & &\\
\hline
(b , v) &-22\degree , -14 &3\degree , 5.8 & 3\degree , 5.8 &1\fdg4 , -2.3 &-17\fdg1 , 8.2 &-3\fdg6 , -3.6\\
points &-31\degree , -14 &-2\degree , 6.1 & -2\degree , 6.1 & -3\degree , 5.8&-17\fdg6 , 11.3 &-6\degree , -7.2\\
for &-35\fdg5 , -6 &-3\degree , 6.1 & -3\degree , 6.1 & -5\fdg5 , 5.9&-19\fdg8 , 12.4 &-10\fdg6 , -12.2\\
upper	  &-42\degree , -7 &-3\fdg5 , 7.2 & -6\degree , 14 & -8\fdg4 , 4.3&-21\fdg2 , 9.3 &-14\fdg1 , -11.8\\
border	  &-61\degree , -6 &-6\degree , 14 & -6\fdg5 , 15.5& -9\degree , 1.7&-23\fdg9 , 4.4 &\\
	  & &-19\degree , 14 & -10\fdg6 , 24.5& -10\degree , 1.4 & \\
	  & &-21\fdg2 , 9.3 & -61\degree, 24.5&-11\degree , 5.3 & &\\	  
	  & &-23\fdg9 , 4.4 & &-13\degree , 3.1 & &\\
	  & & & &-14\fdg6 , 4.2&  &\\
	  & & & &-17\degree , 4.2 & &\\
	  & & & &-18\fdg5 , 6.2 & &\\
	  & & & &-22\fdg1 , 1.3 & &\\
	  & & & &-23\fdg9 , 4.4 & &\\
	  & & & &-25\degree , 6 & &\\
	  & & & &-34\fdg5 , 6& &\\	  
	  & & & &-38\fdg6 , 2.8 & &\\
	  & & & &-61\degree , 7.2 & &\\
	  
\hline
\end{tabular}
\label{tab:CompLim}
\end{table*}

%% file: ResFit.txt
\renewcommand{\arraystretch}{1.7}
\setlength{\tabcolsep}{0.27cm}
\begin{sidewaystable*}
\caption{Best-fit coefficients of the $\gamma$-ray model in each energy band. }
\centering
\begin{tabular}{l|r r r r r|| l | r}
\hline \hline
$\gamma$-ray & \multicolumn{5}{c||}{Energy band [MeV]}& \multicolumn{2}{c}{$\tau_{353}$}\\
model & $10^{2.6}-10^{2.8}$& $10^{2.8}-10^{3.2}$&$10^{3.2}-10^{3.6}$&$10^{3.6}-10^5$&$10^{2.6}-10^5$ & \multicolumn{2}{c}{model}  \\
\hline
$q_{\rm{HI \; Cet}}   $ & 1.13$\pm$0.07$\pm$0.04 & 1.07$\pm$0.05$\pm$0.03 & 1.07$\pm$0.04$\pm$0.04 & 0.99$\pm$0.12$\pm$0.04 & 1.08$\pm$0.04$\pm$0.06 &$y_{\rm{HI \; Cet}}^{\rm a}$ & 1.03$\pm$0.01$\pm$0.02 \\  
$q_{\rm{HI \; TauS}}   $ & 1.10$\pm$0.02$\pm$0.04 & 1.11$\pm$0.02$\pm$0.04 & 1.06$\pm$0.02$\pm$0.04 & 1.05$\pm$0.04$\pm$0.04 & 1.09$\pm$0.01$\pm$0.04 &$y_{\rm{HI \; TauS}}^{\rm a}$ & 1.37$\pm$0.01$\pm$0.04 \\ 
$q_{\rm{HI \; TauN}}   $ & 1.05$\pm$0.03$\pm$0.04 & 1.00$\pm$0.02$\pm$0.03 & 1.02$\pm$0.02$\pm$0.04 & 1.06$\pm$0.05$\pm$0.04 & 1.02$\pm$0.02$\pm$0.05 &$y_{\rm{HI \; TauN}}^{\rm a}$ & 1.10$\pm$0.01$\pm$0.03 \\  
$q_{\rm{HI \; TauM}}   $ & 1.29$\pm$0.03$\pm$0.06 & 1.25$\pm$0.02$\pm$0.06 & 1.21$\pm$0.02$\pm$0.06 & 1.25$\pm$0.05$\pm$0.05 & 1.26$\pm$0.02$\pm$0.06 & $y_{\rm{HI \; TauM}}^{\rm a}$ & 1.42$\pm$0.02$\pm$0.06 \\ 
$q_{\rm{HI \; Cal}}   $ & 1.24$\pm$0.05$\pm$0.02 & 1.20$\pm$0.04$\pm$0.02 & 1.09$\pm$0.04$\pm$0.02 & 1.13$\pm$0.08$\pm$0.02 & 1.15$\pm$0.03$\pm$0.02 & $y_{\rm{HI \; Cal}}^{\rm a}$ & 1.37$\pm$0.02$\pm$0.04 \\ 
$q_{\rm{HI \; Per}}   $ & 1.67$\pm$0.16$\pm$0.06 & 1.41$\pm$0.11$\pm$0.05 & 1.47$\pm$0.12$\pm$0.05 & 1.19$\pm$0.21$\pm$0.04 & 1.42$\pm$0.08$\pm$0.05 & $y_{\rm{HI \; Per}}^{\rm a}$ & 1.65$\pm$0.07$\pm$0.10 \\ 
$q_{\rm{HI \; Gal}}   $ & 0.73$\pm$0.02$\pm$0.03 & 0.75$\pm$0.02$\pm$0.03 & 0.74$\pm$0.02$\pm$0.03 & 0.82$\pm$0.04$\pm$0.04 & 0.74$\pm$0.01$\pm$0.03 & $y_{\rm{HI \; Gal}}^{\rm a}$ & 0.80$\pm$0.04$\pm$0.02 \\ 
$q_{\rm{CO \; Cet}}$ & 1.43$\pm$0.72$\pm$0.23 & 2.50$\pm$0.54$\pm$0.22 & 2.37$\pm$0.56$\pm$0.22 & 0.00$\pm$0.90$\pm$0.00 & 2.09$\pm$0.37$\pm$0.20 & $y_{\rm{CO \; Cet}}^{\rm b}$ & 3.46$\pm$0.17$\pm$0.09 \\ 
$q_{\rm{CO \; TauS}}$ & 2.23$\pm$0.22$\pm$0.09 & 2.35$\pm$0.17$\pm$0.09 & 2.29$\pm$0.18$\pm$0.09 & 1.87$\pm$0.31$\pm$0.10 & 2.26$\pm$0.12$\pm$0.10 & $y_{\rm{CO \; TauS}}^{\rm b}$ & 3.78$\pm$0.09$\pm$0.12 \\ 
$q_{\rm{CO \; TauN}}      $ & 1.90$\pm$0.10$\pm$0.04 & 1.87$\pm$0.08$\pm$0.04 & 1.79$\pm$0.09$\pm$0.04 & 1.86$\pm$0.16$\pm$0.03 & 1.86$\pm$0.06$\pm$0.03 &$y_{\rm{CO \; TauN}}^{\rm b}$ & 3.44$\pm$0.07$\pm$0.04 \\ 
$q_{\rm{CO \; TauM}}   $ & 1.75$\pm$0.06$\pm$0.03 & 1.70$\pm$0.05$\pm$0.03 & 1.62$\pm$0.05$\pm$0.03 & 1.67$\pm$0.09$\pm$0.03 & 1.67$\pm$0.04$\pm$0.04 & $y_{\rm{CO \; TauM}}^{\rm b}$ & 3.58$\pm$0.05$\pm$0.03 \\ 
$q_{\rm{CO \; Cal}}   $ & 2.07$\pm$0.10$\pm$0.04 & 2.09$\pm$0.08$\pm$0.04 & 1.94$\pm$0.09$\pm$0.04 & 1.94$\pm$0.16$\pm$0.05 & 2.04$\pm$0.06$\pm$0.04 & $y_{\rm{CO \; Cal}}^{\rm b}$ & 3.13$\pm$0.08$\pm$0.10 \\ 
$q_{\rm{CO \; Per}}   $ & 1.14$\pm$0.16$\pm$0.13 & 1.35$\pm$0.14$\pm$0.14 & 1.46$\pm$0.13$\pm$0.13 & 1.73$\pm$0.23$\pm$0.14 & 1.38$\pm$0.13$\pm$0.13 & $y_{\rm{CO \; Per}}^{\rm b}$ & 3.34$\pm$0.13$\pm$0.26 \\ 
$q_{\rm{CO \; Gal}}$ & 0.32$\pm$0.25$\pm$0.68 & 0.00$\pm$0.25$\pm$0.43 & 0.05$\pm$0.25$\pm$0.62 & 0.00$\pm$0.25$\pm$0.28 & 0.00$\pm$0.25$\pm$0.48 & $y_{\rm{CO \; Gal}}^{\rm b}$ & 2.91$\pm$0.44$\pm$1.19 \\ 
$q_{\rm{ff} }     $ & 2.96$\pm$0.81$\pm$1.09 & 4.00$\pm$0.77$\pm$0.95 & 2.40$\pm$0.66$\pm$0.86 & 4.20$\pm$1.34$\pm$0.89 & 3.26$\pm$0.57$\pm$0.97 &$y_{\rm{ff} }^{\rm c}     $ & 2.11$\pm$0.61$\pm$1.03 \\ 
$q_{\rm{COsat}}   $ & 4.22$\pm$0.31$\pm$0.10 & 4.44$\pm$0.25$\pm$0.10 & 4.23$\pm$0.26$\pm$0.07 & 4.57$\pm$0.40$\pm$0.09 & 4.40$\pm$0.20$\pm$0.09 &$y_{\rm{COsat}}^{\rm a}   $ & 3.30$\pm$0.10$\pm$0.16 \\ 
$q_{\rm{DNM}}      $ & 6.97$\pm$0.24$\pm$0.07 & 6.83$\pm$0.19$\pm$0.08 & 6.63$\pm$0.21$\pm$0.08 & 7.17$\pm$0.38$\pm$0.09 & 6.86$\pm$0.14$\pm$0.10 &$y_{\rm{DNM}}^{\rm a}$ & 1.52$\pm$0.02$\pm$0.09 \\ 
$q_{\rm{ic}}      $ & 1.40$\pm$0.78$\pm$0.20 & 3.06$\pm$0.72$\pm$0.42 & 2.46$\pm$0.70$\pm$0.26 & 2.31$\pm$0.87$\pm$0.18 & 3.01$\pm$0.48$\pm$0.21 & $y_{\rm{iso}}^{\rm d}$ & -1.54$\pm$0.02$\pm$0.10 \\ 
$q_{\rm{iso}}      $ & 1.96$\pm$0.27$\pm$0.04 & 1.89$\pm$0.30$\pm$0.15 & 1.43$\pm$0.29$\pm$0.06 & 1.85$\pm$0.29$\pm$0.04 & 1.58$\pm$0.16$\pm$0.08 & &\\

\hline \hline
\end{tabular}
\vspace{0.3cm}
\tablefoot{The q coefficients are expressed in $10^{20}$ cm$^{-2}$ (K km s$^{-1}$) for the CO; 3.8 10$^{15}$ cm$^{-2}$Jy$^{-1}$sr for the free-free; $10^{25}$ cm$^{-2}$ for the DNM and CO$_{Sat}$; other q coefficients are simple normalization factors. \\
The y coefficients are expressed as follows: $^{{\rm a}}$ in $10^{-26}$ cm$^{2}$;
$^{{\rm b}}$ in $10^{-6}$ K$^{-1}$km$^{-1}$s;
$^{{\rm c}}$ in 3.8 $10^{-11}$ Jy$^{-1}$sr;
$^{{\rm d}}$ in $10^{-6}$. \\
The first uncertainties are statistical, the second give the standard deviations obtained by varying the \hi spin temperature.
}

 \label{tab:fitscoef}
 \end{sidewaystable*}

%% file: Xco_history.txt
\begin{table*}[!h]
\caption{Estimates of the average $X_{\rm CO}$ factor in units of $10^{20}$ cm$^{-2}$ (K km s$^{-1}$)$^{-1}$. }
\centering
\begin{tabular}{l l l l}
\hline 
$\gamma$-ray telescope & Location & $X_{\rm CO}$ factor & reference\\
\hline
& \multicolumn{2}{c}{Large Galactic scales}  &\\
\hline 
COS-B & 1st quadrant & 2.5$^a$ & \cite{1983ApJ...274..231L} \\
COS-B & Galactic disc & 2.3$^a$ & \cite{1986AeA...154...25B} \\
COS-B & Galactic disc & 2.3 $\pm$ 0.3 & \cite{1988AeA...207....1S} \\
EGRET & Galactic disc & 1.9 $\pm$ 0.2 & \cite{1996AeA...308L..21S} \\
EGRET & Galactic disc & 1.56 $\pm$ 0.05 & \cite{1997ApJ...481..205H} \\
EGRET & 5$^{\circ} < |b| < 80^{\circ}$ & 1.74 $\pm$ 0.03 & \cite{2005Sci...307.1292G} \\
EGRET & Local Arm 3rd quadrant &  1.64 $\pm$ 0.31 & \cite{2001ApJ...555...12D} \\
\textit{Fermi} LAT & Local Arm 2nd quadrant & 1.59 $\pm$ 0.17 & \cite{2010ApJ...710..133A} \\
\textit{Fermi} LAT & Local Arm Cygnus & 1.68 $\pm$ 0.05 & \cite{2012AeA...538A..71A} \\
\textit{Fermi} LAT & Local Arm 3rd quadrant & 2.08 $\pm$ 0.11 & \cite{2011ApJ...726...81A} \\
\textit{Fermi} LAT & between the Local and Perseus arms & 1.93 $\pm$ 0.16 & \cite{2011ApJ...726...81A} \\
\textit{Fermi} LAT & Perseus arm & 1.9 $\pm$ 0.2 & \cite{2010ApJ...710..133A} \\
\hline
& \multicolumn{2}{c}{Nearby clouds} &\\
\hline
COS-B & Oph-Sag & 0.9 $\pm$ 0.4$^a$ & \cite{1984ApJ...281..634L} \\
COS-B & Orion & 2.1 $\pm$ 1.0$^a$ & \cite{1984AeA...139...37B} \\
EGRET & Oph & 1.1 $\pm$ 0.2 & \cite{1994ApJ...436..216H} \\
EGRET & Orion & 1.35 $\pm$ 0.15 & \cite{1999ApJ...520..196D} \\
EGRET & Cepheus & 0.92 $\pm$ 0.14 & \cite{1996ApJ...463..609D} \\
EGRET & Taurus & 1.08 $\pm$ 0.10 & \cite{2001AIPC..587..538D} \\
\textit{Fermi} LAT & Orion & 1.21 $\pm$ 0.02 & \cite{2012ApJ...756....4A} \\
\textit{Fermi} LAT & Cepheus \& Cassiopeia & 0.87 $\pm$ 0.05 & \cite{2010ApJ...710..133A} \\
\textit{Fermi} LAT & Cepheus \& Polaris & 0.63 $\pm$ 0.02 & \cite{2012ApJ...755...22A} \\
\textit{Fermi} LAT & RCrA & 0.99 $\pm$ 0.08 & \cite{2012ApJ...755...22A} \\
\textit{Fermi} LAT & Chamaeleon & 0.69 $\pm$ 0.02 & \cite{2015AeA...582A..31A} \\
\textit{Fermi} LAT & 10$^{\circ} < |b| < 70^{\circ}$ & 0.902 $\pm$ 0.007 & \cite{2015ApJ...806..240C} \\
\hline
\end{tabular}
\tablefoot{$^a$ corrected to absolute CO radiation temperatures according to \cite{1988ApJ...324..248B}}
\label{tab:Xco_history}
\end{table*}